\documentclass[prd,showpacs,amsmath,showkeys,floatfix,amssymb, preprintnumbers, nofootinbib, superscriptaddress]{revtex4} 
\usepackage{hyperref} 
\usepackage{amsmath,amssymb,bm}
\usepackage{epsfig}
\usepackage{graphicx,color}
\usepackage{amsfonts}
\usepackage{epstopdf}
\usepackage{cases}
\usepackage[capitalise]{cleveref}
\usepackage{color}

\usepackage{float}

\newcommand{\nn}{\nonumber}
\newcommand{\beqa}{\begin{eqnarray}}
\newcommand{\eeqa}{\end{eqnarray}}

\newcommand{\bd}[1]{ \mbox{\boldmath $#1$}  }

\begin{document}

\def\ii{\'\i}
\title{BCS solutions and effective quarks energies of the QCD Hamiltonian in the Coulomb
gauge.}

\author{Tochtli Y\'epez-Mart\ii nez}
\email{tochtlicuauhtli.yepez@iems.edu.mx}
\affiliation{Instituto de Educaci\'on Media Superior de la Ciudad de
  M\'exico, Plantel Benito Ju\'arez,    
Av. Zacatl\'an, esq. Cempas\'uchil S/N. Pueblo de San Lorenzo Tezonco,
C.P. 09790, Alcald\ii a Iztapalapa.  Ciudad de M\'exico,  M\'exico.}

\author{Peter O. Hess}
\email{hess@nucleares.unam.mx}
\affiliation{Instituto de Ciencias Nucleares, Universidad Nacional
  Aut\'onoma de M\'exico, Ciudad Universitaria,
Circuito Exterior S/N, A.P. 70-543, 04510 M´exico D.F. Mexico.\\
and\\
Frankfurt Institute for Advanced Studies, J. W. von Goethe University, Hessen, Germany
}

\author{Osvaldo Civitarese}
\email{osvaldo.civitarese@fisica.unlp.edu.ar}
\affiliation{ Departamento de F\'isica, Universidad Nacional de La Plata,
 C.C. 67 (1900), La Plata, Argentina.}

\date{\today}

\begin{abstract}
The exploration of the non-perturbative regime of QCD, that is the
low-energy portion of the hadron spectrum, requires the adoption of
theoretical methods more frequently applied to other, more
conventional, quantum many body systems, like the atomic nucleus,
solid state systems, etc. 
In this work we have adopted, as a first step, 
the well-known BCS method to describe correlations between pairs of quarks
and the associated ground state. 
Going beyond the BCS method would imply 
the inclusion of correlations by means of the TDA or RPA
approximations. Since, we are interested in analyzing the role of
constituent quark-pair correlations in the structure of hadrons we
are restricted to the use of BCS as said before.
The starting Hamiltonian is the
effective Coulomb plus linear potential which we have used in
previous calculations and performed a two-step approach, firstly by
pre-diagonalizing it to built a single particle spectrum, and then,
secondly, by applying the BCS transformations to it. Then, we have
explored the resulting structure of the low energy meson spectra in
terms of quasiparticle degrees of freedom. The dependence of the
results upon the parameters which enter  in the calculations is explored in
detail, at the level of the quasiparticle mean-field approximation.
\end{abstract}

\pacs{12.39.-x, 21.30.Fe, 21.60.Ev, 74.20.Fg}
\keywords{QCD Hamiltonian, Coulomb gauge, BCS method, meson states}
\maketitle

\section{Introduction}
\label{Sec: introduction}

QCD is the favored theory of strong interactions (see
Refs.\cite{Weinberg, Lee-book}). However, due to properties like color
confinement, its low energy regime becomes non-perturbative. In consequence, the description of the observed barionic and mesonic spectrum
cannot be attempted in terms
 of conventional, group symmetry based
methods, without making extreme approximations \cite{SO4-1,SO4-2,SO4-3}. 
This is a paradigmatic situation, e.g: the theory contains the correct degrees of freedom (quarks
and gluons) but the correct description of the observed barionic and
mesonic states seemingly requires going beyond that. 
The effective Hamiltonian of Refs.\cite{ChristLee,Adam1996,Adam2001,Hugo2004,Hugo2011,Yepez2012,Greensite2015,Greensite2016}, which is the QCD
Hamiltonian in the Coulomb gauge, is a good starting point for the application of
many body techniques \cite{Ring, Fetter}. The feasibility of this
approach was shown in previous publications
Refs.\cite{Hess2006,Yepez2010,Arturo2017}. 
A brief resume of the results so far obtained along this line is the following:
In Refs.\cite{Yepez2010,Arturo2017}, we have implemented the
harmonic oscillator solutions 
as a trial basis.
This first approach to treat the effective fermionic sector of the QCD Hamiltonian demonstrates the possibility to include an arbitrary number
of radial and angular excitations, even for the exploration of
hadronic excited states \cite{Arturo2017, Bicudo2016}. We have extended our investigation of the motivated QCD Hamiltonian
based on the framework of the Coulomb gauge, by taking as starting point the effective solutions of quarks and antiquarks,
obtained by diagonalizing the Dirac term in the harmonic oscillator basis \cite{Yepez2010}.
Then, the effects of a confining Coulomb plus linear interaction were taking into account by the implementation
of well known many body techniques like the Tamm-Dancoff-Approximation (TDA) and the  Random-Phase-Approximations
(RPA). Both, the TDA and RPA described meson-like states as collective phonon solutions. In Refs.\cite{Arturo2017}, it was shown
that the lowest energy, meson-like solution of the RPA-method,
describes a highly collective state as a superposition of 
particle-hole states. This picture may be completed by looking at correlations between pairs of quarks which may result in
a superfluid low energy regime . 
Therefore, the observed spectrum of hadrons could be interpreted in terms
of quasiparticle excitations. This is the purpose of the present paper. We proceed gradually by treating the
QCD Hamiltonian in the Coulomb gauge, to construct the effective quark spectrum. 
We will apply the BCS
transformations and solve
the corresponding equations, in order to get the relevant parameters of the model, that is
the occupation factors and gaps. The dependence of the solutions upon
the dimension of the basis and of it in terms 
of coupling schemes is presented and discussed in Section II and Section III, respectively.. The results of the calculations are discussed in
Section \ref{results} and the conclusions are drawn in Section \ref{conclusions}.

\section{From QCD to effective degrees of freedom.}
The implementation of the BCS transformation in a fermionic or bosonic system requires, as a first step, the
definition of a single particle field. In the case of QCD the
literature is rich in the description of such fields \cite{Finger,
  Adler, Yaouanc,Bicudo1990,
Llanes2000,Llanes2002,Nefediev2008,Yepez2010}.
For this we have adopted the Coulomb plus linear QCD Hamiltonian, as it is described next
\subsection{Coulomb gauge QCD Hamiltonian.}
We start from the QCD Hamiltonian
in its canonical Coulomb gauge representation
\cite{ChristLee,Lee-book},
\beqa
{\bd H}^{QCD}
&=&\int \left\{ \frac{1}{2} \left[ \mathcal{J}^{-1}
\mathcal{ {\bf \Pi}}^{tr }  \cdot  \mathcal{J} \mathcal{ {\bf \Pi}}^{tr }
+ {\bf \mathcal{B} }  \cdot  {\bf \mathcal{  B}}\right]
-\overline{{ \psi}}\left(-i{ \gamma}\cdot{ \nabla}+m\right) { \psi}
-g\overline{{ \psi}}{ \gamma}\cdot { A} { \psi}\right\}d {\bf  x}
\nonumber \\
&& +\frac{1}{2}g^{2}\int\mathcal{J}^{-1}\rho^{c}( {\bf  x})
\langle{c,{\bf  x}}|\frac{1}{{ \nabla}\cdot \mathcal{{ D}}}
(-{ \nabla}^{2})\frac{1}{{ \nabla}\cdot \mathcal{{ D}}}
|c^{\prime} {\bf y}\rangle \mathcal{J}\rho^{c^{\prime}}({\bf  y})d {\bf x}d {\bf y}~,
\label{eq1}
\eeqa
which has been widely studied in the past
for the description of several properties of QCD at low energy 
\cite{Adam1996,Adam2001,Zwanziger2003,Greensite2003,Hugo2004,Hugo2011,Peng2008,Yepez2012,Greensite2015,Greensite2016}.
In \cite{Adam2001,Arturo2017} a complete description
of this Hamiltonian has been presented. 
The Hamiltonian of Eq. (\ref{eq1}) includes the relevant interactions between quarks
and gluons through the 
{\it QCD Instantaneous color-Coulomb Interaction}
(QCD-IcCI) between color-charge-densities of quarks and gluons, last
term in Eq. (\ref{eq1}).
At low energy the effects of dynamical gluons in the QCD-IcCI can be represented by the interaction
\begin{equation}\label{eqvxy}
V(| {\bf x} -{\bf y} |)=-\frac{V_C}{| {\bf x} -{\bf y}|}+V_L |{\bf x} -{\bf y}|,
\end{equation}
which is obtained from a self-consistent treatment of the
interaction between color charge-densities \cite{Adam2001,Hugo2004}.

In \cite{Arturo2017}, the Hamiltonian of Eq. (\ref{eq1}) was analyzed when it is restricted
to the quark sector of the theory (no dynamical gluons), and an
effective confining interaction $V(| {\bf x} -{\bf y} |)$ was used to
describe the low energy interaction between color charge
densities. We write
\beqa\label{eq2}
{\bd H}^{QCD}_{eff}
& = & \int\left\{ {\bd \psi}^{\dagger}({\bf x})
(-i{\bd \alpha}\cdot{\bd \nabla}+\beta{m}) {\bd \psi}({\bf x})\right\}d{\bf x}
-\frac{1}{2} \int\rho_{c}({\bf x}) V(|{\bf x}-{\bf y}|)\rho^{c}({\bf y})d{\bf x}d{\bf y}
\nonumber \\
& = & {\bd K} + {\bd H}_{{\rm Coul}}~,
\eeqa
where  $\rho^c({\bf x}) = \psi^\dagger({\bf x}) T^c \psi({\bf x})$
is the quark and antiquark charge density.
In Eq. (\ref{eq2}), the first term is the kinetic energy,
while the second term is the QCD-IcCI in its simplified form.
The fermion field $\psi^\dagger ({\bf x})$,
whose quantization is explained in \cite{Arturo2017},
is expanded in terms of  creation and annihilation
operators in the harmonic oscillator
basis

\beqa\label{fermion-field}
\psi^\dagger ({\bf x}) & = & \sum_{\tau,N l m_l,\sigma C F} 
R^{*}_{Nl}(x) Y^*_{lm_l}(\hat {\bf x}) \chi^\dag_{\sigma} 
~{\bd q}^\dag_{\tau, Nlm_l, \sigma C  F}\nonumber\\
& = & \sum_{Nlm_l, \sigma C F}   R^{*}_{Nl}(x) Y^*_{lm_l}(\hat {\bf x}) \chi^\dag_{\sigma}
\left( {\bd q}^\dag_{\frac{1}{2}, Nlm_l ,\sigma C F} 
+  {\bd q}^\dag_{-\frac{1}{2}, Nlm_l,\sigma C F}\right) ~. 
\label{quant}
\eeqa
with $x=| {\bf x} |$ and  $ R_{Nl}(x)=\textit{N}_{Nl}\exp(-\frac{B_0 x^{2}}{2})
x^{l}L_{\frac{N-l}{2}}^{l+\frac{1}{2}}(B_0 x^{2})$, 
where $L^\lambda_n$ is an associated  Laguerre polynomial
and $( \sqrt{B_0})^{-1}$ is the oscillator length. 
The index $\tau$ denotes  upper ($\tau=\frac{1}{2}$) and lower ($\tau=-\frac{1}{2}$) 
components of the Dirac spinors  in the Dirac-Pauli representation of the Dirac matrices,
and $\sigma,C,F$ denote  spin, color and flavor intrinsic degrees of freedom, respectively. 
The harmonic oscillator basis is chosen, because
it allows to obtain analytic expressions for the
matrix elements of the interaction. 

The diagonalization of the kinetic term (\ref{eq2}) is performed 
using the total spin $ J = l \pm \frac{1}{2}$ representation, 
for a given 
maximal number of quanta $N=N_{{\rm cut}}$, for which we introduce a general transformation
to a basis of effective operators,
\beqa
{\bd q}^\dagger_{\tau (N l)JM_J CF} & = &
\sum_{\lambda \pi k} \left( \alpha^{J,T}_{\tau (Nl),\lambda \pi k}\right)^* 
{\bd Q}^\dagger_{\lambda \pi kJM_J CF}  
~\delta_{\pi,(-1)^{\frac{1}{2}-\tau+l}}
~~~.
\label{eq3}
\eeqa
The index $\lambda = \pm\frac{1}{2}$ refers 
to the pseudo-spin components 
after the diagonalization of the kinetic energy term, and
$k$ runs over all orbital states after  the  diagonalization. 
The value $\lambda = +\frac{1}{2}$ refers to positive energy states
(effective quarks) and the value
$\lambda = -\frac{1}{2}$ to negative energy states (effective
antiquarks), {\it i.e.} 
${\bd Q}^\dagger_{\frac{1}{2} \pi k J M_J CF}  \rightarrow  {\bd
  b}^\dagger_{\pi k J C (Y, T), M_J M_C M_T}$
and
${\bd Q}^\dagger_{-\frac{1}{2} \pi k J M_J CF}  \rightarrow  {\bd
  d}_{\pi  k J C (Y, T), M_J M_C M_T}$.

The transformation coefficients
 $\alpha^{j,T}_{\tau (Nl),\lambda \pi k}$
depend on the type of quarks, whether it is
an up or down quark (equal masses are assumed) or a strange quark ($m_{u,d}<m_s$),
{\it i.e.} from now on we will distinguish flavor representations
according to their  flavor hypercharge and isospin $(Y,T)$, with  magnetic projection $M_T$.
For the transformation coefficients and the matrix elements of the kinetic energy term of the Hamiltonian
only the dependence on the flavor isospin is given,
because the flavor-hypercharge is fixed by $T$.

The eigenvalue problem to be solved acquires the following form
\beqa\label{eq:prediag}
\sum_{\tau_i  N_i l_i } 
 \alpha^{j,T}_{\tau_1,(N_1 l_1),   \lambda_1 \pi_1 k_1} 
K^{j,T}_{\tau_1(N_1l_1),\tau_2(N_2l_2)}
 \alpha^{j,T}_{\tau_2,(N_2 l_2), \lambda_2 \pi_2 k_2} 
=\varepsilon_{\lambda_1 \pi_1 k_1 J C (Y, T)}
~\delta_{\lambda_1 \lambda_2}\delta_{\pi_1 \pi_2}\delta_{k_1 k_2}~,
\eeqa
where we have  taken the transformation coefficients of Eq.\ (\ref{eq3}) to be real.

The correct identification
of the effective quark and antiquark degrees of freedom  is one of the most important
steps in order to describe hadronic states. By implementing the
harmonic oscillator basis to treat the fermionic sector of the QCD
Hamiltonian, we have performed a prediagonalization and identified
effective quarks and antiquarks as a linear combinations of the bare
quarks and antiquarks. In terms of the effective quarks and antiquarks,
the kinetic energy term acquires the following structure
\beqa
K=\sum_{k \pi \gamma} \varepsilon_{k\pi \gamma}\sum_{\mu}
\left(
\bd{ b}^\dag_{k\pi\gamma\mu}
\bd{ b}^{k\pi\gamma\mu}
-
\bd{ d}^{k\pi\gamma\mu}
\bd{ d}^\dag_{k\pi\gamma\mu}
\right)~,
\eeqa
where the creation and annihilation operators of the effective quarks
and antiquarks are given by $\bd{ b}^\dag_{k\pi\gamma\mu},~\bd{
  b}^{k\pi\gamma\mu}$ and $\bd{ d}^\dag_{k\pi\gamma\mu},~\bd{  d}^{k\pi\gamma\mu}$,
respectively.
The upper and lower indices indicate the principal number
($k=1,2,\dots$) which run over all orbital states and
the parity ($\pi=\pm$) while $\gamma,~\mu$ are short hand notation for the
particle spin, color and flavor hypercharge and isospin
representations  $\gamma=\{J,C,(Y,T)\}$
and their magnetic projections $\mu=\{M_J, M_C, M_T\}$, respectively. The flavor
hypercharge and isospin quantum numbers for quarks are given by
$(Y,T)=(\frac{1}{3},\frac{1}{2}),~(-\frac{2}{3},0)$. The quarks and
antiquarks belong to a triplet $C=(10)$ and anti-triplet $\bar{C}=(01)$
color irreducible representations (irreps), respectively, which are
conjugate representations.


The QCD-IcCI term, in its simplified form $({\bd H}_{{\rm Coul}})$,
rewritten in terms of effective quarks and antiquarks operators is
given by \cite{Arturo2017}
\beqa
\bd{H}_{Coul} &=& - \frac{1}{2} \sum_{L }
\sum_{ \lambda_i {\bf q}_i }
V^{L}_{ \{\lambda_i {\bf q}_i \} }
\Big(
\left[\mathcal{F}_{\lambda_1 {\bf q}_1, \lambda_2 {\bf q}_2;\gamma_{f_0}}
\mathcal{F}_{\lambda_3 {\bf q}_3, \lambda_4 {\bf q}_4;\bar \gamma_{f_0}}\right]^{\gamma_0}_{\mu_0}
+\left[\mathcal{F}_{\lambda_1 {\bf q}_1, \lambda_2 {\bf q}_2;\gamma_{f_0}}
\mathcal{G}_{\lambda_3 {\bf q}_3, \lambda_4 {\bf q}_4;\bar \gamma_{f_0}}\right]^{\gamma_0}_{\mu_0}\nn\\
&~&~~~~~~~~~~~~~~~~~~~~~
+\left[\mathcal{G}_{\lambda_1 {\bf q}_1, \lambda_2 {\bf q}_2;\gamma_{f_0}}
\mathcal{F}_{\lambda_3 {\bf q}_3, \lambda_4 {\bf q}_4;\bar \gamma_{f_0}}\right]^{\gamma_0}_{\mu_0}
+\left[\mathcal{G}_{\lambda_1 {\bf q}_1, \lambda_2 {\bf q}_2;\gamma_{f_0}}
\mathcal{G}_{\lambda_3 {\bf q}_3, \lambda_4 {\bf q}_4;\bar \gamma_{f_0}}\right]^{\gamma_0}_{\mu_0}
\Big)~,
\label{eq13}
\eeqa
where we have compacted the single particle orbital number, parity and
irreps into the short-hand notation ${\bf q}_i=k_i \pi_{q_i} \gamma_{q_i} $,
and use for the (flavorless) quantum numbers of the intermediate coupling in the
interaction $\gamma_{f_0}= \{L(11)(0,0)\}$ and for
their magnetic projections $\mu_{f_0}=\{M_L,M_C,0\}$.
The conjugate representations satisfy $\bar \gamma_{f_0}=\gamma_{f_0}$ and
$\bar \mu_{f_0}=\{-M_L,\bar M_C,0\}$. 
For the total
couplings (upper index) and magnetic numbers (lower index) of the
interaction, we have used  $\gamma_0=\{0,(00),(0,0)\}$
and $\mu_0=\{0,0,0\}$ respectively.
The operators $\mathcal{F}$ and $\mathcal{G}$ are given by
\beqa
\mathcal{F}_{\lambda_1 {\bf q}_1, \lambda_2 {\bf q}_2;\gamma_{f_0}, \mu_{f_0}}
&=&
\frac{1}{\sqrt{2}} \bigg\{ \delta_{\lambda_1,\frac{1}{2}} \delta_{\lambda_2,\frac{1}{2}}
\left[
{\bd b}^\dagger_{ {\bf q}_1 }  \otimes  {\bd b}_{  \bar {\bf q}_2 }
\right]^{\gamma_{f_0}}_{\mu_{f_0}}
-
\delta_{\lambda_1,-\frac{1}{2}} \delta_{\lambda_2,-\frac{1}{2}}
\left[ {\bd d}_{ \bar {\bf q}_1 }   \otimes   {\bd d}^\dagger_{ \bar {\bf q}_2 }
\right]^{\gamma_{f_0}}_{\mu_{f_0}} \bigg\}
\nonumber\\
\mathcal{G}_{\lambda_1 {\bf q}_1, \lambda_2 {\bf q}_2;\gamma_{f_0},  \mu_{f_0}}
&=& \frac{1}{\sqrt{2}}  \bigg\{
\delta_{\lambda_1,-\frac{1}{2}} \delta_{\lambda_2,\frac{1}{2}}
\left[   {\bd d}_{ {\bf q}_1 }   \otimes  {\bd b}_{  \bar {\bf q}_2 }
\right]^{\gamma_{f_0}}_{\mu_{f_0}}
-
\delta_{\lambda_1,\frac{1}{2}} \delta_{\lambda_2,-\frac{1}{2}}
\left[ {\bd b}^\dagger_{ {\bf q}_1 }  \otimes  {\bd d}^\dagger_{  \bar {\bf q}_2}
\right]^{\gamma_{f_0}}_{\mu_{f_0}}    \bigg\}
~~.
\label{eq12}
\eeqa

In this basis, and using the above introduced states, the matrix elements of the interaction are given by 
\beqa\label{eq-new-matrix-elements}
V^{L}_{ \{\lambda_i \pi_i k_i  J_i Y_i T_i \} }
&=&\sum_{\tau_i N_i l_i}~ V_{\{N_i l_i J_i\}}^{L}~
\alpha^{J_1,T_1}_{\tau_1(N_1l_1),\lambda_1,\pi_1,k_1}
\alpha^{J_2,T_2}_{\tau_2(N_2l_2),\lambda_2,\pi_2,k_2}
\alpha^{J_3,T_3}_{\tau_3(N_3l_3),\lambda_3,\pi_3,k_3}
\alpha^{J_4,T_4}_{\tau_4(N_4l_4),\lambda_4,\pi_4,k_4} \nonumber\\
&\times&
~\delta_{\tau_1\tau_2} \delta_{\tau_3\tau_4}~
\delta_{ \pi_1 , (-1)^{\frac{1}{2}-\tau_1 + l_1} }
\delta_{  \pi_2 , (-1)^{\frac{1}{2}-\tau_2 + l_2} }
\delta_{ \pi_3 , (-1)^{\frac{1}{2}-\tau_3 + l_3} }
\delta_{  \pi_4 , (-1)^{\frac{1}{2}-\tau_4 + l_4} } \nonumber\\
& \times&
(-1)^{\frac{1}{3}+\frac{Y_1}{2}+T_1}\frac{\sqrt{2T_1+1}}{\sqrt{3}}\delta_{T_2T_1}\delta_{Y_2 Y_1}
~(-1)^{\frac{1}{3}+\frac{Y_3}{2}+T_3}\frac{\sqrt{2T_3+1}}{\sqrt{3}}\delta_{T_4T_3}\delta_{Y_4 Y_3}~.
\eeqa
The matrix elements in the harmonic oscillator basis ($V_{\{N_i l_i J_i\}}^{L}$) are analytic and
actually easy to compute.

\subsection{Bogoliubov Transformation .}
Here, we apply a canonical transformation from the particle
(antiparticle) quark basis to another basis known as the quasiparticle
basis, in order to approximately diagonalize part of the QCD
interaction. The method is well known in many-particle
physics under the name {\it Bogoliubov Transformation} \cite{Ring, Fetter}.

The use of the Bogoliubov transformations in fermionic and bosonic
systems,  to account for pairing-type of correlations and the
consequences of it in terms of the building of ground state
correlations, both in finite and continuos system, has been documented
in hundred (if not thousand) of paper since they have been proposed in
the earlier $1950'$s. A compilation of such a references, for the case
of nuclear systems, may be found in \cite{50yearsBCS}.
In the case of hadronic physics the notions related to the use of pairs
correlations can be found in the chapter 21 of the book by
S. Weinberg  \cite{Weinberg}, where the concept of spontaneous symmetry
breaking has been explicitly applied to interacting hadrons.
Another relevant reference about the use of the BCS approach is
\cite{Bes}.

The transformation from the effective quark degrees of freedom to
the quasiquark basis is done by means of the Bogoliubov
transformations. Herewith we shall apply the transformations for each quark-flavor, separately. Thus, the reference (ground state) state will be referred to as a quark condensate with a definite flavor. In terms of the BCS 
solutions, which we are going to introduce below, this amounts 
to the construction of a set of parameters (occupation numbers and gap) for each flavor. Then, the Hamiltonian of Eq.(\ref{eq2}) is written in terms
of quasi-quark operators and the standard conditions of the BCS
theory are applied to it by asking the one-quasiparticle term to be
diagonal, the pairs terms to vanish and by extracting from the
constant terms the corresponding gaps. These steps are shown next.
The creation and annihilation quasi-quark operators are 
written as
\beqa
{\bd B}^\dag_{k_i\pi_i, \gamma_i \mu_i}
&=& u_{k_i\pi_i, \gamma_i }{\bd b}^\dag_{k_i\pi_i ,\gamma_i \mu_i}
-v_{k_i\pi_i, \gamma_i}{\bd d}_{k_i\pi_i ,\gamma_i \mu_i}\nonumber\\
{\bd D}^{\dag k_i\pi_i, \gamma_i \mu_i}
 &=& u_{k_i\pi_i, \gamma_i }{\bd  d}^{\dag  k_i\pi_i, \gamma_i \mu_i}
+v_{k_i\pi_i, \gamma_i }{\bd b}^{k_i\pi_i, \gamma_i \mu_i}
\eeqa
for the creation operators, and
\beqa
{\bd B}^{k_i\pi_i, \gamma_i \mu_i}
&=& u^*_{k_i\pi_i, \gamma_i }{\bd b}^{k_i\pi_i, \gamma_i \mu_i}
-v^*_{k_i\pi_i, \gamma_i }{\bd d}^{\dag k_i\pi_i, \gamma_i \mu_i}\nonumber\\
{\bd D}_{k_i\pi_i, \gamma_i \mu_i} 
&=& u^*_{k_i\pi_i, \gamma_i } {\bd d}_{k_i\pi_i,  \gamma_i \mu_i} 
+v^*_{k_i\pi_i, \gamma_i } {\bd b}^\dag_{k_i\pi_i, \gamma_i \mu_i} 
\eeqa 
for the annihilation ones, respectively. In
the above equations we have used a short hand notation to denote the
states, that is:  $\gamma_i=\{J_i C_i (Y_i,T_i)\}$ and
$\mu_i=\{M_{J_i} M_{C_i} M_{T_i}\}$. The coefficients $u$ and $v$
should be taken as real.

The inverse transformations are 
given by
\beqa\label{BCS-trans}
{\bd b}^\dag_{k_i\pi_i, \gamma_i \mu_i}
&=& u^*_{k_i\pi_i, \gamma_i }{\bd B}^\dag_{k_i\pi_i, \gamma_i \mu_i}
+v_{k_i\pi_i, \gamma_i }{\bd D}_{k_i\pi_i, \gamma_i \mu_i}\nonumber\\
{\bd d}^{\dag  k_i\pi_i, \gamma_i \mu_i}
&=& u^*_{k_i\pi_i, \gamma_i }{\bd  D}^{\dag  k_i\pi_i, \gamma_i \mu_i}
-v_{k_i\pi_i, \gamma_i }{\bd B}^{k_i\pi_i, \gamma_i \mu_i}
\eeqa

A crucial step  in the treatment, leading to the transformation between ordinary particles (in this case fermions like the quarks and antiquarks) 
to quasiparticles is the replacement of the ordinary vacuum $|0 \rangle$ by  the BCS vacuum $|BCS \rangle$, which amounts to a spontaneous symmetry breaking, which is expressed by the weak identity:   
\beqa \label{gap}
\langle BCS| {\bd b}^\dagger_k {\bd d}^{\dagger k}    |BCS\rangle \sim \Delta_k \ne 0
\eeqa
where $ \Delta_k$ is the state (flavor) dependent gap.
At the same time the following conditions:  
\beqa
{\bd B}^k |BCS\rangle=0
\eeqa
should be obeyed.
As an additional comment about the meaning of the BCS vacuum
expectation value Eq. (\ref{gap}),  it is worth to mention that it
plays the role of a mass, since the square of the gap will appear in
the definition of the quasiparticle energies. This is indeed the case
of up and down quarks, because for these flavors the gap is
non-zero. For strange quarks the gap is always zero, as we shall show in
Section IV.B.

The method implemented to determine the actual value of the parameters $u$ and $v$ of Eq.(\ref{BCS-trans}), for each flavor, is a variational one, where the terms of the Hamiltonian to be varied are kept up to the forth power of these parameters. The convergence of the solutions was tested as a function of the maximum number of quanta $N_{cut}$.

\section{Quasiparticle Hamiltonian}
The terms of the transformed Hamiltonian are obtained by replacing in Eq. (\ref{eq2}), 
the tensorial product of quarks and antiquark operators by their quasiparticle  expression. 
From  Eq.(\ref{BCS-trans}) we get for the terms entering Eq. (\ref{eq12})
\beqa
&&\left[ {\bd b}^\dagger_{ {\bf q}_1 }  \otimes  {\bd b}_{  \bar {\bf      q}_2 }\right]^{\gamma_{f_0}}_{\mu_{f_0}}\nonumber\\
&&=\langle  \gamma_{1} \mu_{1}, \bar {\gamma}_{2} \bar {\mu}_{2} |\gamma_{f_0} \mu_{f_0} \rangle 
(-1)^{\gamma_{2}-\mu_{2}}
{\bd b}^\dagger_{ k_1 \pi_1 \gamma_{1} \mu_{1}}  {\bd b}^{ k_2 \pi_2 \gamma_{2} \mu_{2} }\nonumber\\
&&=\langle  \gamma_{1} \mu_{1}, \bar {\gamma}_{2} \bar {\mu}_{2} |\gamma_{f_0} \mu_{f_0} \rangle 
(-1)^{\gamma_{2}-\mu_{2}}
\left(  u^*_{ k_1 \pi_1 \gamma_{1} }{\bd B}^\dag_{ k_1 \pi_1 \gamma_{1} \mu_{1}}
+v_{ k_1 \pi_1 \gamma_{1} }{\bd D}_{ k_1 \pi_1 \gamma_{1} \mu_{1} }  \right)
\left(  u_{  k_2 \pi_2 \gamma_{2} }{\bd B}^ {  k_2 \pi_2 \gamma_{2} \mu_2 }
+v^*_{ k_2 \pi_2 \gamma_{2} }{\bd D}^{\dag  k_2 \pi_2 \gamma_{2} \mu_2 }  \right)
\nonumber\\
&&
\left[ {\bd d}_{ {\bf q}_1}    \otimes   {\bd d}^\dagger_{ \bar {\bf q}_2 }\right]^{\gamma_{f_0}}_{\mu_{f_0}}\nonumber\\
&&=\langle  \gamma_{1} \mu_{1}, \bar {\gamma}_{2} \bar {\mu}_{2} |\gamma_{f_0} \mu_{f_0} \rangle 
(-1)^{\gamma_{2}+\mu_{2}}
{\bd d}_{ k_1 \pi_1 \gamma_{1} \mu_{1}}  {\bd d}^{\dagger k_2 \pi_2 \gamma_{2} \mu_{2} }\nonumber\\
&&=\langle  \gamma_{1} \mu_{1}, \bar {\gamma}_{2} \bar {\mu}_{2} |\gamma_{f_0} \mu_{f_0} \rangle 
(-1)^{\gamma_{2}+\mu_{2}}
\left(  u_{ k_1 \pi_1 \gamma_{1} }{\bd D}_{ k_1 \pi_1 \gamma_{1} \mu_{1}}
-v^*_{ k_1 \pi_1 \gamma_{1} }{\bd B}^\dagger_{ k_1 \pi_1 \gamma_{1} \mu_{1} }  \right)
\left(  u^*_{  k_2 \pi_2 \gamma_{2} }{\bd D}^ {\dagger  k_2 \pi_2 \gamma_{2} \mu_2 }
-v_{ k_2 \pi_2 \gamma_{2} }{\bd B}^{  k_2 \pi_2 \gamma_{2} \mu_2 }  \right)
\nonumber\\
&&
\left[   {\bd d}_{ {\bf q}_1 }   \otimes  {\bd b}_{  \bar {\bf q}_2 }
\right]^{\gamma_{f_0}}_{\mu_{f_0}}\nonumber\\
&&=\langle  \gamma_{1} \mu_{1}, \bar {\gamma}_{2} \bar {\mu}_{2} |\gamma_{f_0} \mu_{f_0} \rangle 
(-1)^{\gamma_{2}-\mu_{2}}
{\bd d}_{ k_1 \pi_1 \gamma_{1} \mu_{1}}  {\bd b}^{ k_2 \pi_2 \gamma_{2} \mu_{2} }\nonumber\\
&&=\langle  \gamma_{1} \mu_{1}, \bar {\gamma}_{2} \bar {\mu}_{2} |\gamma_{f_0} \mu_{f_0} \rangle 
(-1)^{\gamma_{2}-\mu_{2}}
\left(  u_{ k_1 \pi_1 \gamma_{1} }{\bd D}_{ k_1 \pi_1 \gamma_{1} \mu_{1}}
-v^*_{ k_1 \pi_1 \gamma_{1} }{\bd B}^\dagger_{ k_1 \pi_1 \gamma_{1} \mu_{1} }  \right)
\left(  u_{  k_2 \pi_2 \gamma_{2} }{\bd B}^ {  k_2 \pi_2 \gamma_{2} \mu_2 }
+v^*_{ k_2 \pi_2 \gamma_{2} }{\bd D}^{\dagger  k_2 \pi_2 \gamma_{2} \mu_2 }  \right)\nonumber\\
&&
\left[ {\bd b}^\dagger_{ {\bf q}_1 }  \otimes  {\bd d}^\dagger_{  \bar {\bf q}_2}
\right]^{\gamma_{f_0}}_{\mu_{f_0}}\nonumber\\
&&=\langle  \gamma_{1} \mu_{1}, \bar {\gamma}_{2} \bar {\mu}_{2} |\gamma_{f_0} \mu_{f_0} \rangle 
(-1)^{\gamma_{2}+\mu_{2}}
{\bd b}^\dagger_{ k_1 \pi_1 \gamma_{1} \mu_{1}} {\bd d}^{\dagger k_2 \pi_2 \gamma_{2} \mu_{2} }\nonumber\\
&&=\langle  \gamma_{1} \mu_{1}, \bar {\gamma}_{2} \bar {\mu}_{2} |\gamma_{f_0} \mu_{f_0} \rangle 
(-1)^{\gamma_{2}+\mu_{2}}
\left(  u^*_{ k_1 \pi_1 \gamma_{1} }{\bd B}^\dag_{ k_1 \pi_1 \gamma_{1} \mu_{1}}
+v_{ k_1 \pi_1 \gamma_{1} }{\bd D}_{ k_1 \pi_1 \gamma_{1} \mu_{1} }\right)
\left(  u^*_{  k_2 \pi_2 \gamma_{2} }{\bd D}^ {\dagger  k_2 \pi_2 \gamma_{2} \mu_2 }
-v_{ k_2 \pi_2 \gamma_{2} }{\bd B}^{  k_2 \pi_2 \gamma_{2} \mu_2 }  \right)\nonumber\\
\eeqa
and similarly for the kinetic energy terms of
Eq. (\ref{eq:prediag}). 
We have used the short-hand notation
$\langle  \gamma_{1} \mu_{1}, \bar {\gamma}_{2} \bar {\mu}_{2} |\gamma_{f_0} \mu_{f_0} \rangle $
and  $(-1)^{\gamma_2 \pm \mu_2}$
for the product of the spin, color and isospin-flavor Clebsch–Gordan
 coefficients and phases \cite{Arturo2017}, respectively.

The next step in
our derivation consists of taking normal order  respect to the quasiparticle  vacuum 
and collecting the different contributions to the 
Hamiltonian with constant terms $H_{00}$, one creation-one
annihilations terms $H_{11}$,  two-quasiparticle terms $H_{20}$ and $H_{02}$. 
The value of the gap is extracted, for each channel, by solving the
set of BCS equations (see Section \ref{BCS-eqs}) and
the remanent of the transformed Hamiltonian may be  treated, as
explain before, in the TDA or in the RPA basis (\cite{Arturo2017}) 
to
describe correlations between pairs of quasiparticles \cite{TDA-RPA-2021}. 
The terms of the transformed Hamiltonian, which are relevant 
to determine the extend of the
superfluid correlations, are the following:

i) Constant term ${H}_{00}$
\beqa\label{H00}
\tilde{H}_{00}&=&\sum_{k\pi\gamma}\tilde {\varepsilon }_{k\pi \gamma}
\left( 2 v_{k\pi\gamma}^2-1\right) \Omega_{k\pi\gamma}
-\sum_{k_i, \pi_i \gamma_i} h_{00}(k_i,\pi_i,\gamma_i)
\eeqa

ii) One quasiparticle-term ${H}_{11}$
\beqa\label{H11}
\tilde{H}_{11}
&=& \sum_{k\pi \gamma }\tilde{\varepsilon}_{k\pi, \gamma_{q} }
(u_{k\pi, \gamma}^2-v_{k\pi, \gamma}^2)
({\bd B}^\dag_{k\pi, \gamma } \cdot{\bd B}^{ k\pi, \gamma}
+{\bd D}^{\dag  k\pi, \gamma} \cdot {\bd  D}_{ k\pi,  \gamma })
\nonumber\\
&-& 
\sum_{k_i \pi_i \gamma_i}
\Big\{
h_{11}( k_i,\pi_i,\gamma_i)
{\bd B}^\dag_{k_1\pi_1, \gamma_{1} } \cdot
{\bd B}^{ k_4\pi_1, \gamma_{1} }              
+
h_{11}(k_i,\pi_i,\gamma_i)
{\bd D}^{\dag k_4\pi_1, \gamma_{1}} \cdot
{\bd D}_{ k_1\pi_1, \gamma_{1} }
\Big\} 
\nonumber\\            
&-& 
\sum_{k_i \pi_i \gamma_i}
\Big\{
h_{11}(k_i,\pi_i,\gamma_i)
{\bd B}^\dag_{k_3\pi_3, \gamma_{3} } \cdot
{\bd B}^{ k_2\pi_3, \gamma_{3} } 
+
h_{11}(k_i,\pi_i,\gamma_i)
{\bd D}^{\dag k_2\pi_2, \gamma_{2}} \cdot
{\bd D}_{ k_3\pi_2, \gamma_{2} }
\Big\}
\nonumber\\
 \eeqa

iii) Two-quasiparticle terms: $H_{20}+H_{02}$
\beqa\label{H20}
H_{20}+H_{02}&=& \sum_{k\pi \gamma} 2 \tilde{\varepsilon}_{k\pi, \gamma }
 u_{k\pi, \gamma}  v_{k\pi, \gamma}
( {\bd B}^\dag_{k\pi, \gamma} \cdot{\bd D}^{\dag k\pi, \gamma}
+{\bd D}_{k\pi, \gamma } \cdot{\bd B}^{ k\pi, \gamma }  )        \nonumber\\
&-& 
\sum_{ k_i \pi_i\gamma_i}
\Big\{
h_{20}(k_i,\pi_i,\gamma_i)
{\bd B}^\dag_{k_1\pi_1, \gamma_{1} }  \cdot {\bd D}^{\dag k_4\pi_1, \gamma_{1} }
+
h_{02}(k_i,\pi_i,\gamma_i)
{\bd D}_{k_1\pi_1, \gamma_{1} }  \cdot  {\bd B}^{ k_4\pi_1, \gamma_{1} }
\Big\} 
\nonumber\\
&-& 
\sum_{k_i\pi_i\gamma_i}
\Big\{
h_{20}(k_i,\pi_i,\gamma_i)
{\bd B}^\dag_{k_3\pi_3, \gamma_{3} } \cdot {\bd D}^{\dag k_2\pi_3, \gamma_{3} }  
+
h_{02}(k_i,\pi_i,\gamma_i)
{\bd D}_{k_3\pi_3, \gamma_{3} } \cdot {\bd B}^{ k_2\pi_3, \gamma_{3} }
\Big\}
\nonumber\\
 \eeqa

The explicit form of the coefficients  $h_{ab}(k_i,\pi_i,\gamma_i)$ of the previous equations is given in the Appendix A, for each term of the transformed Hamiltonian.

\subsection{BCS equations.}
\label{BCS-eqs}
The terms $H_{11}$ and $ H_{20}+H_{02}$ of Eqs. (18) and (19) can be ordered in terms of the following variables 
\beqa
X_{k_1}&=&u_{k_1}^2-v_{k_1}^2\nonumber\\
Y_{k_1}&=&2u_{k_1}v_{k_1}\nonumber\\ 
\eeqa
which depend only on the quasiparticle operator indices, $k_i$. Here, we are
using a short-hand notation $k_i=k_i \pi_i, \gamma_i$. The
interaction terms are also ordered in terms of the structures
$\big( u_{k_2}^2-v_{k_2}^2 \big)$ and 
$\big( u_{k_2}v_{k_2}\big)$,  
being $k_2$ an internal index.  
The parameters $u_k$ and $v_k$ are determined self-consistently.

Notice that the resulting ordered Hamiltonian displays terms of the
type 
\beqa\label{structures}
&&\big( u_{k_2}^2-v_{k_2}^2 \big) X_{k_1}\nonumber\\
&&\big( u_{k_2}^2-v_{k_2}^2 \big) Y_{k_1}\nonumber\\
\eeqa
and also
\beqa
&&\big( u_{k_2}v_{k_2}\big) X_{k_1}\nonumber\\
&&\big( u_{k_2}v_{k_2}\big) Y_{k_1} ~.
\eeqa
When the sum on the internal indeces is performed explicitly this
structure decouples as explained in the Appendix \ref{int-terms},
leading to the equations  
\beqa\label{Eq1-BCS}
\Sigma_{k_1} X_{k_1}+\Delta_{k_1} Y_{k_1} =E_{k_1} 
\eeqa
\beqa\label{Eq2-BCS}
-\Delta_{k_1} X_{k_1} +\Sigma_{k_1} Y_{k_1}=0
\eeqa
where $\Sigma_{k_1}$, $\Delta_{k_1}$
and the quasiparticle energy $E_{k_1}$ are given by
\beqa\label{sigma_and_X}
\Sigma_{k_1}&=& \varepsilon_{k_1}
+\bar{V}^\Sigma_{k_1k_2k_2k_1}(u_{k_2}^2-v_{k_2}^2) \\
\nonumber\\
\Delta_{k_1}&=&\bar{V}^\Delta_{k_1k_2k_2k_1}   (u_{k_2}v_{k_2})
\nonumber\\
E_{k_1}&=&\sqrt{\Sigma_{k_1}^2 +\Delta_{k_1}^2}
\eeqa
with
\beqa
&&\bar{V}^\Sigma_{k_1\pi_1J_1Y_1T_1,
~~k_2\pi_2J_2Y_2T_2 ,
~~k_2\pi_2J_2Y_2T_2,
~~ k_1\pi_1J_1Y_1T_1}=\nonumber\\
&&
- \frac{1}{2} ~
\sum_{L}  \sum_{\lambda_i}
~\left(\frac{1}{2}\right)\frac{\sqrt{8(2L+1)}}{9} \frac{   (-1)^{L+J_2-J_1} }{ 2J_1+1 }\nonumber\\
&&\times
\bigg\{
\sum_{\tau_i N_i l_i}~ V_{\{N_i l_i J_i\}}^{L}
\alpha^{J_1,T_1}_{\tau_1(N_1l_1),\lambda_1,\pi_1,k_1}
\alpha^{J_2,T_2}_{\tau_2(N_2l_2),\lambda_2,\pi_2,k_2}
\alpha^{J_3,T_3}_{\tau_3(N_3l_3),\lambda_3,\pi_3,k_3}
\alpha^{J_4,T_4}_{\tau_4(N_4l_4),\lambda_4,\pi_4,k_4} \nonumber\\
&&\times
~\delta_{\tau_1\tau_2} \delta_{\tau_3\tau_4}~
\delta_{ \pi_1 , (-1)^{\frac{1}{2}-\tau_1 + l_1} }
\delta_{  \pi_2 , (-1)^{\frac{1}{2}-\tau_2 + l_2} }
\delta_{ \pi_3 , (-1)^{\frac{1}{2}-\tau_3 + l_3} }
\delta_{  \pi_4 , (-1)^{\frac{1}{2}-\tau_4 + l_4} }      \bigg\}\nonumber\\
&&\times
( \delta_{\pi_1 \pi_4}) (\delta_{k_2 k_3}\delta_{\pi_2 \pi_3})
( \delta_{J_2 J_3}\delta_{J_1 J_4} )
( \delta_{T_1 T_2}\delta_{T_2 T_3} \delta_{T_3  T_4})
( \delta_{Y_1 Y_2}\delta_{Y_2 Y_3} \delta_{Y_3  Y_4})\nonumber\\
&&\times 
(\delta_{\lambda_1+\frac{1}{2}}\delta_{\lambda_2+\frac{1}{2}}
\delta_{\lambda_3+\frac{1}{2}}\delta_{\lambda_4+\frac{1}{2}}
-\delta_{\lambda_1-\frac{1}{2}}\delta_{\lambda_2+\frac{1}{2}}
\delta_{\lambda_3+\frac{1}{2}}\delta_{\lambda_4-\frac{1}{2}})_{\mbox{average}}
\eeqa
and
\beqa
&&\bar{V}^\Delta_{k_1\pi_1J_1Y_1T_1,
~~k_2\pi_2J_2Y_2T_2 ,
~~k_2\pi_2J_2Y_2T_2,
~~ k_1\pi_1J_1Y_1T_1}=\nonumber\\
&&
- \frac{1}{2} ~
\sum_{L}  \sum_{\lambda_i}
~\left(\frac{1}{2}\right)\frac{\sqrt{8(2L+1)}}{9} \frac{   (-1)^{L+J_2-J_1} }{ 2J_1+1 }\nonumber\\
&&\times
\bigg\{
\sum_{\tau_i N_i l_i}~ V_{\{N_i l_i J_i\}}^{L}
\alpha^{J_1,T_1}_{\tau_1(N_1l_1),\lambda_1,\pi_1,k_1}
\alpha^{J_2,T_2}_{\tau_2(N_2l_2),\lambda_2,\pi_2,k_2}
\alpha^{J_3,T_3}_{\tau_3(N_3l_3),\lambda_3,\pi_3,k_3}
\alpha^{J_4,T_4}_{\tau_4(N_4l_4),\lambda_4,\pi_4,k_4} \nonumber\\
&&\times
~\delta_{\tau_1\tau_2} \delta_{\tau_3\tau_4}~
\delta_{ \pi_1 , (-1)^{\frac{1}{2}-\tau_1 + l_1} }
\delta_{  \pi_2 , (-1)^{\frac{1}{2}-\tau_2 + l_2} }
\delta_{ \pi_3 , (-1)^{\frac{1}{2}-\tau_3 + l_3} }
\delta_{  \pi_4 , (-1)^{\frac{1}{2}-\tau_4 + l_4} }      \bigg\}\nonumber\\
&&\times
( \delta_{\pi_1 \pi_4}) (\delta_{k_2 k_3}\delta_{\pi_2 \pi_3})
( \delta_{J_2 J_3}\delta_{J_1 J_4} )
( \delta_{T_1 T_2}\delta_{T_2 T_3} \delta_{T_3  T_4})
( \delta_{Y_1 Y_2}\delta_{Y_2 Y_3} \delta_{Y_3  Y_4})\nonumber\\
&&\times 2
(\delta_{\lambda_1+\frac{1}{2}}\delta_{\lambda_2+\frac{1}{2}}
\delta_{\lambda_3-\frac{1}{2}}\delta_{\lambda_4-\frac{1}{2}}
+\delta_{\lambda_1-\frac{1}{2}}\delta_{\lambda_2+\frac{1}{2}}
\delta_{\lambda_3-\frac{1}{2}}\delta_{\lambda_4+\frac{1}{2}})~,
\eeqa
where the summation on the internal indices is performed.

These non-linear equations (\ref{Eq1-BCS}) and (\ref{Eq2-BCS})  are then solved for each of the
quark-flavors and they are known as state depended
BCS-equations \cite{Ring,Fetter}, because $\Sigma_{k_1}$ and $\Delta_{k_1}$ depend on the flavor.

The numerical analysis of the matrix elements shows that in the $H_{1 1}$ term, the matrix elements associated with the structures 
$\big( u_{k_2}^2-v_{k_2}^2 \big) Y_{k_1}$
and 
$\big( u_{k_2}v_{k_2}\big) X_{k_1}$ are very small compared to the
matrix elements associated to the structures 
$\big( u_{k_2}^2-v_{k_2}^2 \big) X_{k_1}$
and 
$\big( u_{k_2}v_{k_2}\big) Y_{k_1}$. 
On the other hand, the numerical analysis of the matrix elements shows that in the $H_{02}$ and $H_{20}$ terms, the matrix elements associated with the structures 
$\big( u_{k_2}^2-v_{k_2}^2 \big) X_{k_1}$
and 
$\big( u_{k_2}v_{k_2}\big) Y_{k_1}$
are very small compared to the
matrix elements associated to the structures 
$\big( u_{k_2}v_{k_2}\big) X_{k_1}$
and .
$\big( u_{k_2}^2-v_{k_2}^2 \big) Y_{k_1}$. 

The procedure to obtain the solutions of Eqs. (\ref{Eq1-BCS}) and (\ref{Eq2-BCS}) consist of the
variation of the quantities $u_{k_i}$ and $v_{k_i}$, such that the
iteration stop when the correlation energy $E_{k_1}$ reaches stability.

After introducing these expressions we are in
conditions to present and discuss the results of our calculations


\section{Results and Discussions.}\label{results}

In this section we shall present the results of our calculations. We have started by:
i) studying the effects associated to the renormalization of the parameters entering the definition of the interaction (Coulomb plus linear potential), and, 
ii) by comparing the spectra resulting from the diagonalization of the interaction with the quasiparticle spectrum, as a function of the cut-off parameter ($N_{cut}$) which gives the size of the radial basis,  
Next, we have constructed the spectra for meson-like states as
uncorrelated two-quasiquark states, for different values of total
isospin T. The theoretical spectra include states up to 2 GeV and they
are compared to the experimental ones \cite{PDG2020}, for different values of the
angular momentum and parity $J^{\pi}$. The density of states is shown
as a function of $N_{cut}$.It is worth mentioning that the main
purpose of this work is to discuss how feasible is the application of the BCS formalism to treat the non-perturbative regime of QCD. The main aspect of the comparison between calculations and data will focus on the density of states, that is to say we shall investigate if the space of uncorrelated two-quasiparticle states  is dense enough to establish connections with data.

\subsection{Renormalization of the interaction and masses.}

To absorb any dependence of the configurational space on the cutoff
$N_{cut}$, we have to implement a renormalization procedure.
A cutoff in the number of oscillator quanta $N_{cut}$ is introduced
to perform the numerical calculations. Such a truncation is similar to a
momentum cutoff regularization. However, instead of a continuous cutoff that truncates
the integrals in momentum space, our cutoff is discrete and truncates the space
of admissible oscillations.
The aim of the renormalization procedure is to keep the eigenvalues of
the BCS equations $E_{k_i}$ unchanged, and hence the masses of the physical
states, invariant under changes of the cutoff $N_{cut}$. Unfortunately, the exact implementation
of such a procedure is very difficult due to the nonlinear dependence of
the eigenvalues on the parameters that appear in the Hamiltonian, the bare quark
masses  $m_{u,d}$ and $m_s$ and the couplings $V_C$ and $V_L$. 
Here, we present the renormalization results which, in fact, does succeed in keeping the low-energy meson-like
spectrum approximately cutoff-independent.

\begin{figure}[H]
\centering
\includegraphics[width=0.7\textwidth]{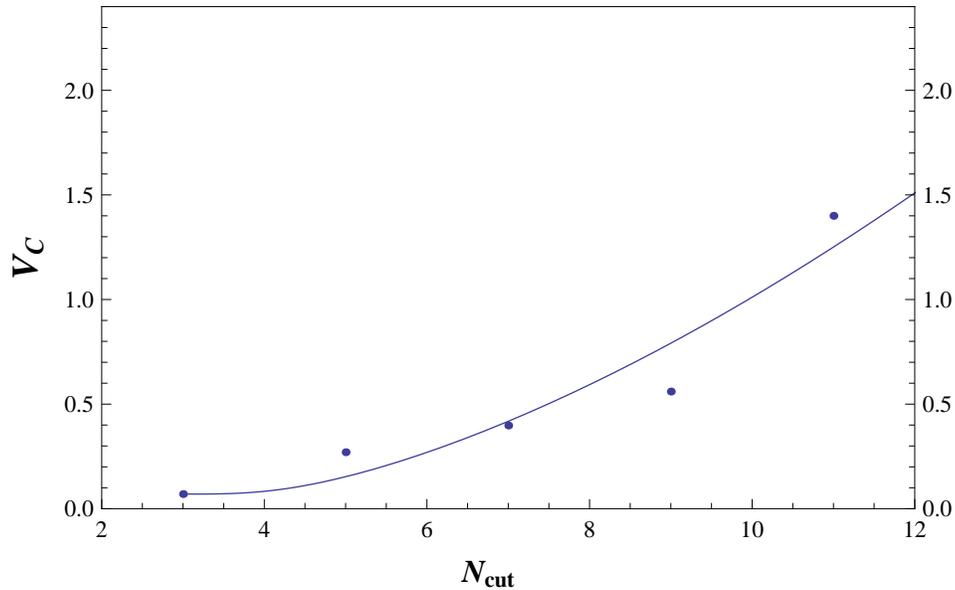}
\caption{Renormalization of the Coulomb interaction: dependence of the parameter $V_C$ upon the size of the radial basis ($N_{cut}$). The actual values are represented by dots, the line is to guide the eye. We are using natural units alltrough the text.}
\label{figalpha}
\end{figure}

\begin{figure}[H]
\centering
\includegraphics[width=0.7\textwidth]{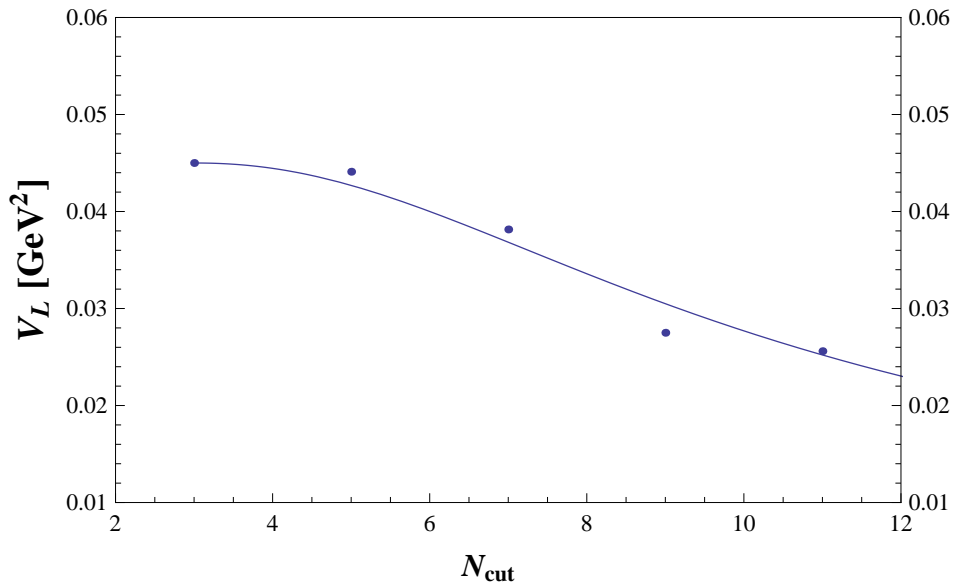}
\caption{Renormalization of the linear interaction:dependence of the parameter $V_L$ upon the size of the radial basis ($N_{cut}$). The actual values are represented by dots, the line is to guide the eye.} \label{figbeta}
\end{figure}

We shall proceed by studying the dependence of the parameters which
enter into the definition of the interaction, which are $V_C$ and $V_L$ of Eq.(\ref{eqvxy}) . Figs.\ref{figalpha}
and \ref{figbeta} show the dependence of the parameters $V_C$ and $V_L$ upon the value of the
cut-off, $N_{cut}$, of the radial basis. In doing so we have varied both $V_C$ and $V_L$ such that the resulting value of the gap remains constant at the level of approximately 0.2 GeV.

\begin{figure}[H]
\centering
\includegraphics[width=0.7\textwidth]{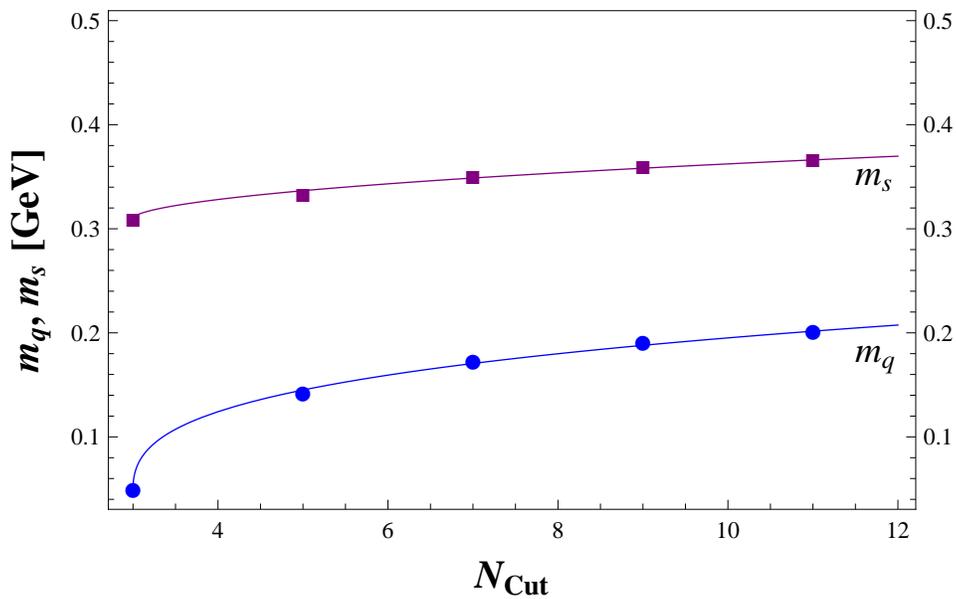}
\caption{Renormalization of the quark masses 
as a function of the cut-off parameter $N_{cut}$, for up and 
down quarks (dots) and strange quarks (squares), 
the continuous line serves to guide the eye.
}  \label{figmasses}
\end{figure}

\subsection{Quasiparticle spectrum.}

\begin{figure}[H]
\centering
\includegraphics[width=0.7\textwidth]{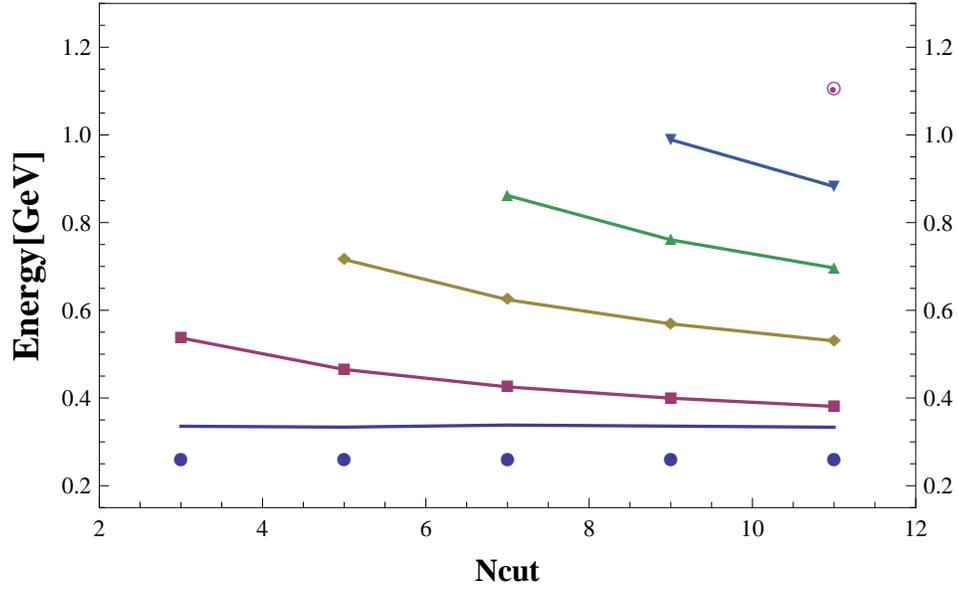}
\caption{Prediagonalization energies (symbols) for $T=\frac{1}{2}$ states
and
  Quasiparticle (solid lines) energies versus $N_{cut}$.}
\label{prediag}
\end{figure}

\begin{figure}[H]
\centering
\includegraphics[width=0.7\textwidth]{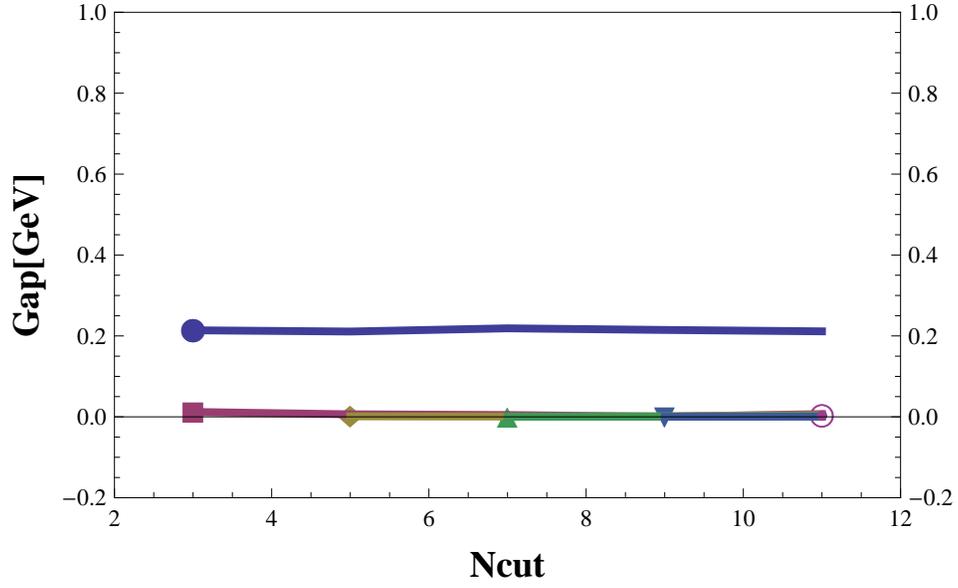}
\caption{Gap solutions for quarks up and down vs $N_{cut}$.}
\label{upgap}
\end{figure}

Proceeding in the same manner, with  the couplings of the previous
subsection, we have diagonalized  the one quasi-particle sector of the
Hamiltonian Eqs. (\ref{H00})-(\ref{H20}) and solved the BCS equations
(\ref{Eq1-BCS}) and (\ref{Eq2-BCS}).
The results are shown in Figures \ref{prediag} and \ref{upgap}

\begin{figure}[H]
\centering
\includegraphics[width=0.7\textwidth]{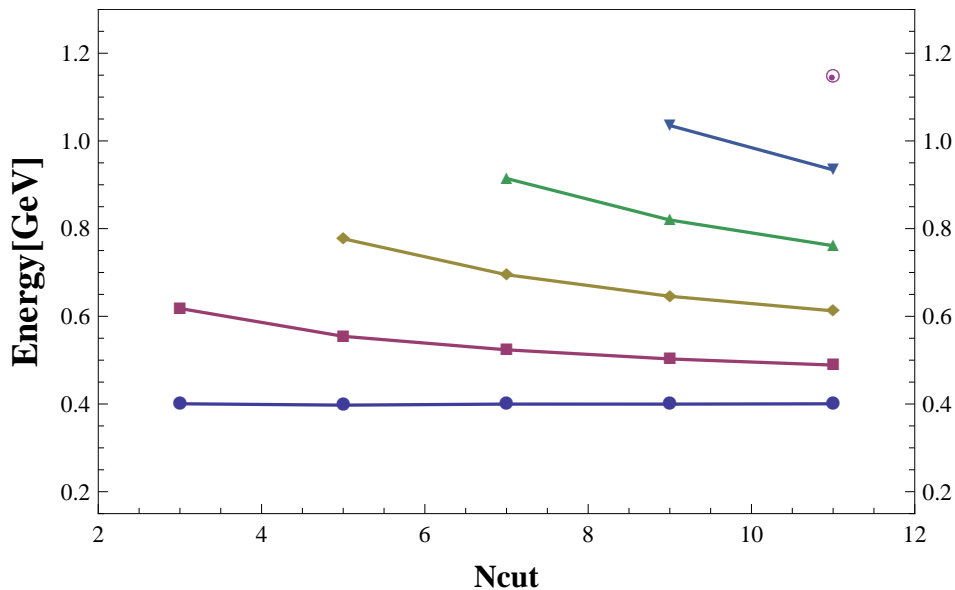}
\caption{Prediagonalization energies (symbols) 
and
  Quasiparticle (solid lines) energies, for $T=\frac{1}{2}$, versus $N_{cut}$, for the strange-quark sector.}
\label{presbcs}
\end{figure}

The same sort of results, for the case of the strange quarks, show that they are insensitive to pairing correlations and their gap is null, see Figure (\ref{presbcs})

From the comparison between the energies obtained by diagonalizing the one particle sector of the Hamiltonian and those corresponding to the quasiparticles, we may conclude that 
in both cases the spectrum reaches a harmonic limit for large values of $N_{cut}$. The constancy of the gap for the up and down quarks is well illustrated by the results shown in Figures 4 and 5, where $\Delta \approx$ 0.2 GeV.

\subsubsection{Meson $B^\dag D^\dag$ spectrum as a function of $N_{cut}$}
\begin{figure}[H]
\centering
\includegraphics[width=0.7\textwidth]{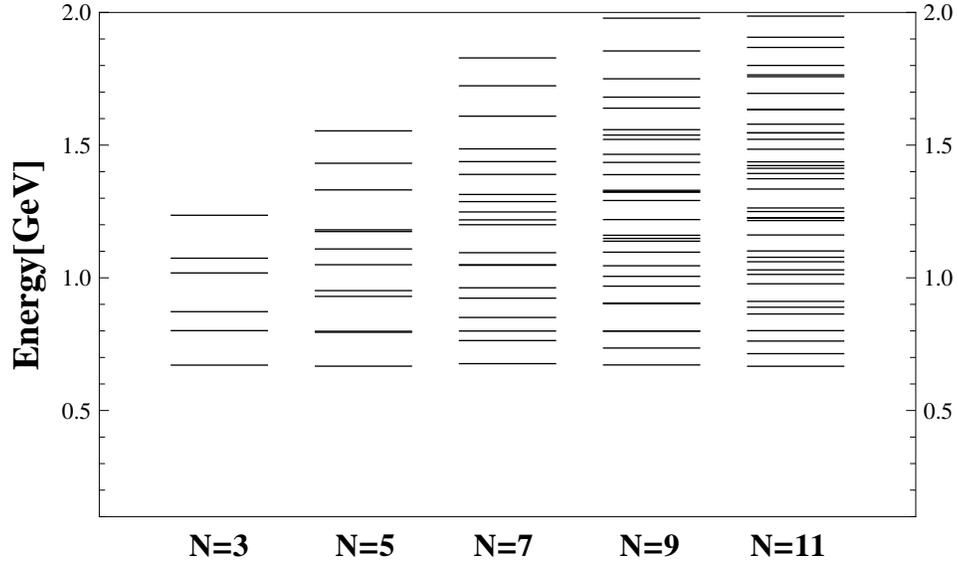}
\caption{Two-Quasiparticle meson-like  spectrum, for pairs of up and down quasi-quarks, isospin $T=0,1$ states,  vs $N_{cut}$.}
\label{mesonsupd}
\end{figure}

\begin{figure}[H]
\centering
\includegraphics[width=0.7\textwidth]{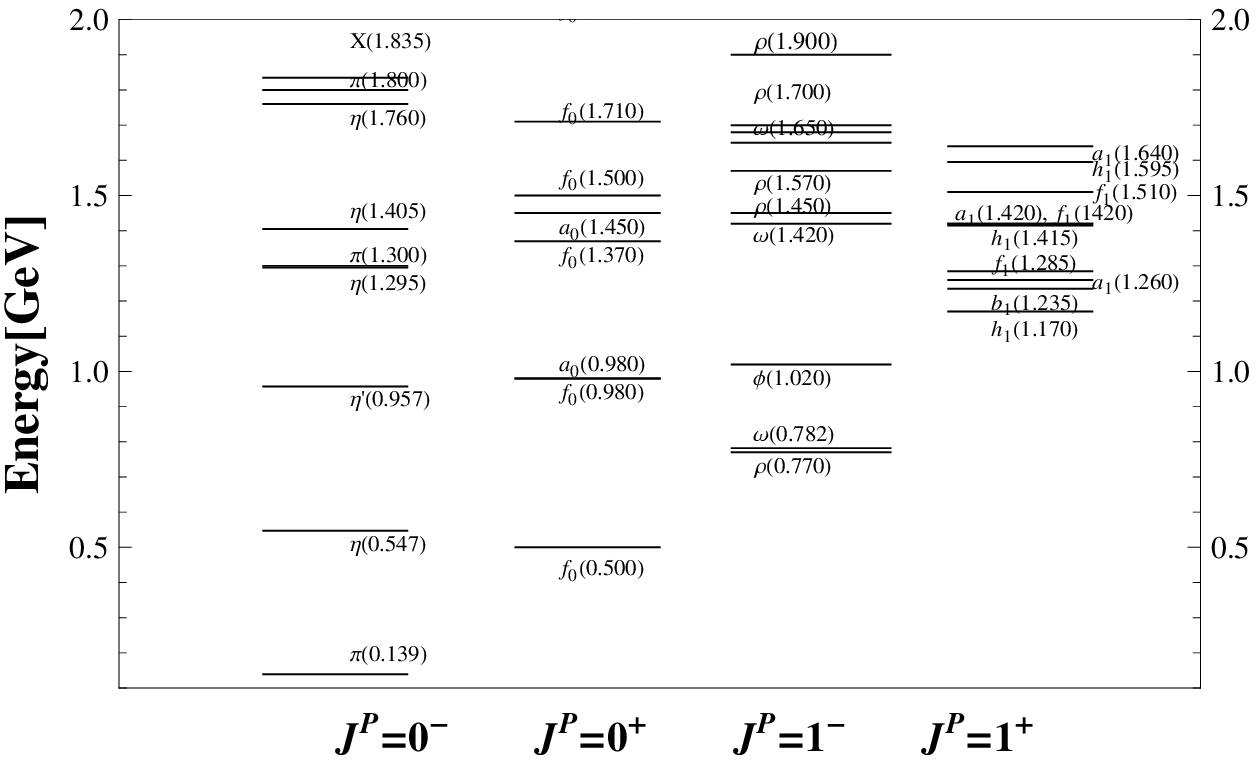}
\caption{Experimental meson spectrum for isospin $T=0,1$, up to 2 Gev.}
\label{ExpMesons}
\end{figure}

\begin{figure}[H]
\centering
\includegraphics[width=0.7\textwidth]{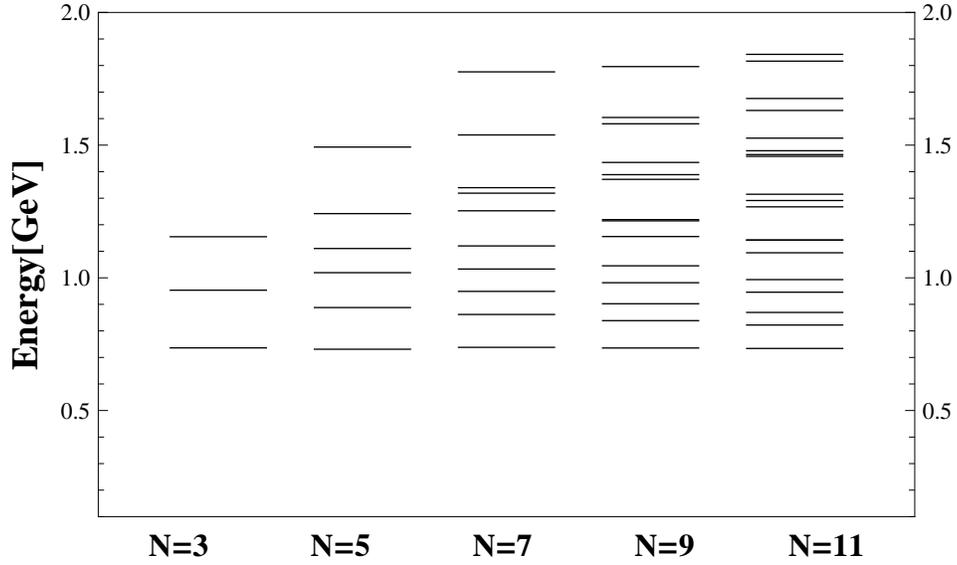}
\caption{Quasiparticle meson spectrum, for pairs of up/down and strange quasi-quarks, isospin $T=1/2$ meson states, vs $N_{cut}$}
\label{strangem}
\end{figure}

\begin{figure}[H]
\centering
\includegraphics[width=0.7\textwidth]{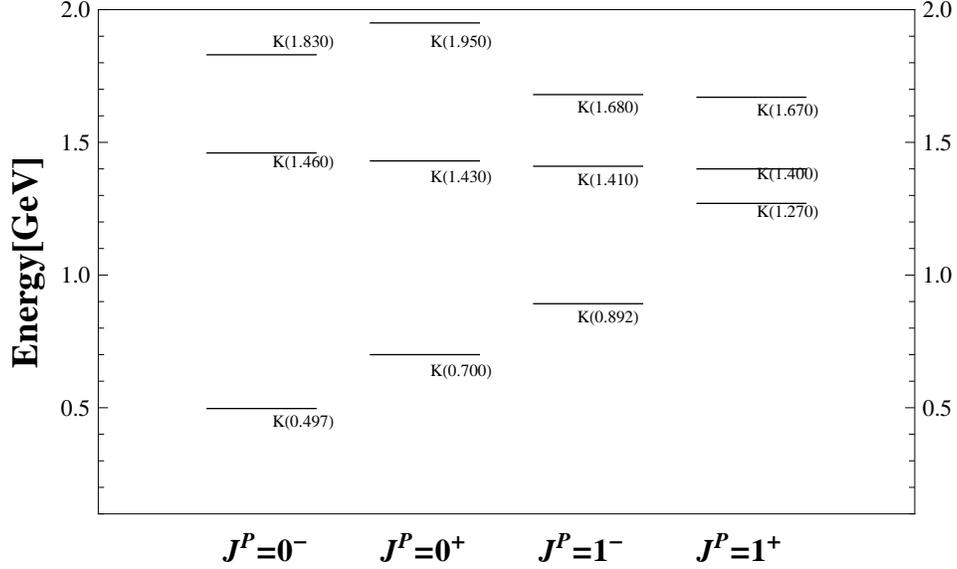}
\caption{Experimental meson spectrum for isospin $T=1/2$, up to 2 Gev.}
\label{ExpMesonsK}
\end{figure}

Meson states, of positive and negative parities, are described as
two-quasiparticle states. In Figure \ref{mesonsupd} the
spectrum of two-quasiparticles, for the subspace $T=0,1$, is shown
as a function of $N_{cut}$. The density of states increases as the number
of states in the basis increases. It is seen that the spacing of
levels is not regular and that for some energies the spectrum
becomes nearly degenerate, a feature which is also observed
experimentally, as shown in Figures \ref{ExpMesons} and
\ref{ExpMesonsK}. The two-quasiparticle spectrum shows a breaking of accidental degenerancies for larger values of $N_{cut}$.
The theoretical and experimental results, for the strange sector, are shown in Figures \ref{strangem} and \ref{ExpMesonsK}, respectively.

Considering that the theoretical results have been
obtained at the quasiparticle level, that is without including
residual interactions between pairs of quasiparticles, the overall
tendency of them 
follows that of the experiment. 
This is particularly
true for the sector of medium and high energies. This is encouraging
because the addition of the residual terms of the
interaction between pairs of quasiparticles, when treated in the
context of non-perturbative linearization methods, like the TDA and
RPA approaches, could certainly improve the agreement, as it was the
case of the schematic forces used in Ref.\cite{Arturo2017}.

In figure \ref{mesonsupd}, we show the dependence of the calculated density of
states as a function of the cutoff. The figure is not meant to be
compared with data but rather show the gross  features of the
spectrum. It is observed that in the low energy portion of the
spectrum, the states are arrange in groups of levels with gaps between
the groups.

The overall features of the theoretical results may be summarized in the following.

(a) The effect of the Coulomb interaction is minor compared to the linear term whose dependence upon  the dimension of the basis is larger.

(b) The calculated spectra show a sort of pairs-like  grouping of levels, a behavior which seems to be confirm by the experimental data.

(c) The experimental spectra show larger spacing between  group  of levels, but this feature is also shown, although at a smaller scale, by the theoretical results.

(d) The theoretical spectra saturate, for larger values of the dimension of the basis.

\section{Conclusions.}\label{conclusions}

In this work we have treated the Coulomb plus linear QCD Hamiltonian
by applying non-perturbative techniques which originate in other
branches of physics. 
The procedure was based on the transformation
from the quark to the quasiquark basis by applying BCS
transformations. 
A renormalization of the mass and interaction parameters was
performed.
The stability of the results was tested by
increasing the dimension  of the radial basis used in the
calculations. 
We have calculated the gaps and  quasiparticle
energies. The two-quasiparticle configurations, that is  mesons
and kaons like states, even 
at this level of approximation show
features similar to those exhibited by the experiments. 
It is expected that going beyond the BCS approximation, by including
interactions between pairs of quasiparticles, would allow for a more
detailed correspondence between theoretical and experimental
results, as done in preliminary studies using schematic models \cite{SO4-1,SO4-2,Arturo2017}.
Work is in progress about the use of the TDA and RPA
methods in this context.


\section*{Acknowledgments}

P.O.H. acknowledges financial support from PAPIIT-DGAPA
(IN100421). O.C acknowledges the support of the CONICET 
and of the ANPCyT of Argentina (PIP-616).


\appendix
\section{Interaction Terms.}
\label{int-terms}

In this Appendix we are given the explicit expressions of the
coefficients of the transformed  Hamiltonian Eqs. (17-19).

To begin with, we write the coefficients of the constant term
$\hat{H}_{00}$  Eq. (\ref{H00}) , which is:  
\beqa
h_{00}(k_i,\pi_i,\gamma_i)
&=&\frac{1}{2}\sum_{L}\sum_{\lambda_i}
~\left(\frac{1}{2}\right) \bigg\{ \bigg\}
\sqrt{8} \sqrt{2L+1}  (-1)^{L+J_2-J_1} \nonumber\\
&\times&
( \delta_{\pi_1 \pi_4}) ( \delta_{\pi_2 \pi_3}) (\delta_{k_1 k_4} (\delta_{k_2 k_3}
( \delta_{J_2 J_3}\delta_{J_1 J_4} )
( \delta_{T_1 T_2}\delta_{T_2 T_3} \delta_{T_3  T_4} \delta_{T_4 T_1})
( \delta_{Y_1 Y_2}\delta_{Y_2 Y_3} \delta_{Y_3  Y_4} \delta_{Y_4 Y_1})
\nonumber\\
&\times\Big\{&
+ (u_{k_2\pi_2, \gamma_{2}}^2) ~\times~
\big[
+(\delta_{-++-}) u_{k_1\pi_1, \gamma_{1} }^2
+(\delta_{++++}) v_{k_1\pi_1, \gamma_{1} }^2 \nonumber\\
&&
~~~~~~~~~~~~~~~~~~~~~~~~~~
+(\delta_{-+++}) (u_{k_1\pi_1, \gamma_{1} }v_{k_1\pi_1, \gamma_{1} })
+(\delta_{+++-}) (v_{k_1\pi_1, \gamma_{1} }u_{k_1\pi_1, \gamma_{1} })
\big]
\nonumber\\
%
%
&&+ (v_{k_2\pi_2, \gamma_{2}}^2) ~\times~
\big[
+(\delta_{----}) u_{k_1\pi_1, \gamma_{1} }^2
+(\delta_{+--+}) v_{k_1\pi_1, \gamma_{1} }^2 \nonumber\\
&&
~~~~~~~~~~~~~~~~~~~~~~~~~~
+(\delta_{---+}) (u_{k_1\pi_1, \gamma_{1} }v_{k_1\pi_1, \gamma_{1} })
+(\delta_{+---}) (v_{k_1\pi_1, \gamma_{1} }u_{k_1\pi_1, \gamma_{1} })
\big]
\nonumber\\
%
%
&&+ (u_{k_2\pi_2, \gamma_{2}} v_{k_2\pi_2, \gamma_{2}}) ~\times~
\big[
-(\delta_{-+--}) u_{k_1\pi_1, \gamma_{1} }^2
-(\delta_{++-+}) v_{k_1\pi_1, \gamma_{1} }^2 \nonumber\\
&&
~~~~~~~~~~~~~~~~~~~~~~~~~~
-(\delta_{-+-+}) (u_{k_1\pi_1, \gamma_{1} }v_{k_1\pi_1, \gamma_{1} })
-(\delta_{++--}) (v_{k_1\pi_1, \gamma_{1} }u_{k_1\pi_1, \gamma_{1} })
\big]
\nonumber\\
%
%
&&+ (v_{k_2\pi_2, \gamma_{2}} u_{k_2\pi_2, \gamma_{2}}) ~\times~
\big[
-(\delta_{--+-}) u_{k_1\pi_1, \gamma_{1} }^2
-(\delta_{+-++}) v_{k_1\pi_1, \gamma_{1} }^2 \nonumber\\
&&
~~~~~~~~~~~~~~~~~~~~~~~~~~
-(\delta_{--++}) (u_{k_1\pi_1, \gamma_{1} }v_{k_1\pi_1, \gamma_{1} })
-(\delta_{+-+-}) (v_{k_1\pi_1, \gamma_{1} }u_{k_1\pi_1, \gamma_{1} })
~~~~\Big\}
\eeqa
with
\beqa\label{Ap-brackets}
\bigg\{ \bigg\}&=&\sum_{\tau_i N_i l_i}~ V_{\{N_i l_i J_i\}}^{L}
\alpha^{J_1,T_1}_{\tau_1(N_1l_1),\lambda_1,\pi_1,k_1}
\alpha^{J_2,T_2}_{\tau_2(N_2l_2),\lambda_2,\pi_2,k_2}
\alpha^{J_3,T_3}_{\tau_3(N_3l_3),\lambda_3,\pi_3,k_3}
\alpha^{J_4,T_4}_{\tau_4(N_4l_4),\lambda_4,\pi_4,k_4} \nonumber\\
&\times&
~\delta_{\tau_1\tau_2} \delta_{\tau_3\tau_4}~
\delta_{ \pi_1 , (-1)^{\frac{1}{2}-\tau_1 + l_1} }
\delta_{  \pi_2 , (-1)^{\frac{1}{2}-\tau_2 + l_2} }
\delta_{ \pi_3 , (-1)^{\frac{1}{2}-\tau_3 + l_3} }
\delta_{  \pi_4 , (-1)^{\frac{1}{2}-\tau_4 + l_4} } \nonumber\\
\eeqa
and where $\Omega_{k\pi\gamma} =\sum_{\mu} 1=\sum_{m_J c f} 1$ is the
degeneracy  of the state $k\pi\gamma$. 
The $\delta_{\pm \pm \pm \pm}$ terms are a short hand for
\beqa\label{Ap-deltas}
\delta_{\lambda_1, \pm \frac{1}{2}}\delta_{\lambda_2, \pm \frac{1}{2}}
\delta_{\lambda_3, \pm \frac{1}{2}}\delta_{\lambda_4, \pm \frac{1}{2}}
\eeqa

Similarly, the four terms which appear in the definition of
$\hat{H}_{11}$, Eq. (\ref{H11})
are:
coefficient in front of $B^\dag_{k_1} B^{k_4}$ 
\beqa
&&h_{11}( k_i,\pi_i,\gamma_i)\nonumber\\
&&=
\frac{1}{2} ~
\sum_{L}  \sum_{\lambda_i}
~\left(\frac{1}{2}\right)\frac{\sqrt{8(2L+1)}}{9} \frac{   (-1)^{L+J_2-J_1} }{ 2J_1+1 }
\bigg\{ \bigg\}
( \delta_{\pi_1 \pi_4}) (\delta_{k_2 k_3}\delta_{\pi_2 \pi_3})
( \delta_{J_2 J_3}\delta_{J_1 J_4} )
( \delta_{T_1 T_2}\delta_{T_2 T_3} \delta_{T_3  T_4})
( \delta_{Y_1 Y_2}\delta_{Y_2 Y_3} \delta_{Y_3  Y_4})\nonumber\\
&&\times
\big\{ (u_{k_2\pi_2, \gamma_2}^2) ~\times~
\big[
(\delta_{++++}) (u_{k_1\pi_1, \gamma_{1} }u_{k_4\pi_1,\gamma_{1} })
-(\delta_{+++-}) (u_{k_1\pi_1, \gamma_{1}}v_{k_4\pi_1,\gamma_{1} })
\nonumber\\
&&
~~~~~~~~~~~~~~~~~~~~~~~~~~~~~
-(\delta_{-+++}) (v_{k_1\pi_1, \gamma_{1} }u_{k_4\pi_1,\gamma_{1} })
+(\delta_{-++-}) (v_{k_1\pi_1, \gamma_{1} }v_{k_4\pi_1, \gamma_{1} })
\big]
\nonumber\\
&&- (v_{k_2\pi_2, \gamma_{2}}^2) ~\times~
\big[
-(\delta_{----}) (v_{k_1\pi_1, \gamma_{1} }v_{k_4\pi_1, \gamma_{1} })
+(\delta_{---+}) (v_{k_1\pi_1, \gamma_{1}}u_{k_4\pi_1,\gamma_{1} })
\nonumber\\
&&
~~~~~~~~~~~~~~~~~~~~~~~~~
+(\delta_{+---}) (u_{k_1\pi_1, \gamma_{1} }v_{k_4\pi_1,\gamma_{1} })
-(\delta_{+--+}) (u_{k_1\pi_1, \gamma_{1}}u_{k_4\pi_1,\gamma_{1} })
\big]
\nonumber\\
&&+ (u_{k_2\pi_2, \gamma_{2}} v_{k_2\pi_2, \gamma_{2}}) ~\times~
\big[
(\delta_{++--}) (u_{k_1\pi_1,  \gamma_{1}}v_{k_4\pi_1,\gamma_{1} })
-(\delta_{++-+}) (u_{k_1\pi_1, \gamma_{1}}u_{k_4\pi_1,\gamma_{1} })
\nonumber\\
&&
~~~~~~~~~~~~~~~~~~~~~~~~~~~~~~~~~
-(\delta_{-+--}) (v_{k_1\pi_1, \gamma_{1} }v_{k_4\pi_1,\gamma_{1} })
+(\delta_{-+-+}) (v_{k_1\pi_1, \gamma_{1} }u_{k_4\pi_1,  \gamma_{1} })
\big]
\nonumber\\
&&+ (v_{k_2\pi_2, \gamma_{2}} u_{k_2\pi_2, \gamma_{2}}) ~\times~
\big[
(\delta_{--++}) (v_{k_1\pi_1,  \gamma_{1}}u_{k_4\pi_1,\gamma_{1} })
-(\delta_{--+-}) (v_{k_1\pi_1, \gamma_{1}}v_{k_4\pi_1,\gamma_{1} })
\nonumber\\
&&
~~~~~~~~~~~~~~~~~~~~~~~~~~~~~~~~~
-(\delta_{+-++}) (u_{k_1\pi_1, \gamma_{1}}u_{k_4\pi_1,\gamma_{1} })
+(\delta_{+-+-}) (u_{k_1\pi_1, \gamma_{1} }v_{k_4\pi_1, \gamma_{1} })
\big]
\big\}~,\nonumber\\
\eeqa
coefficient in front of $D^{\dag k_4} D_{k_1}$ is
\beqa
&&h_{11}(k_i,\pi_i,\gamma_i) \nonumber\\
&&=
\frac{1}{2} ~
\sum_{L}  \sum_{\lambda_i}
~\left(\frac{1}{2}\right)\frac{\sqrt{8(2L+1)}}{9} \frac{   (-1)^{L+J_2-J_1} }{ 2J_1+1 }
\bigg\{ \bigg\}
( \delta_{\pi_1 \pi_4}) (\delta_{k_2 k_3}\delta_{\pi_2 \pi_3})
( \delta_{J_2 J_3}\delta_{J_1 J_4} )
( \delta_{T_1 T_2}\delta_{T_2 T_3} \delta_{T_3  T_4})
( \delta_{Y_1 Y_2}\delta_{Y_2 Y_3} \delta_{Y_3  Y_4})\nonumber\\
&&\times
\big\{ (u_{k_2\pi_2, \gamma_{2}}^2) ~\times~
\big[
-(\delta_{++++}) (v_{k_1\pi_1, \gamma_{1} }v_{k_4\pi_1, \gamma_{1} })
-(\delta_{+++-}) (v_{k_1\pi_1,\gamma_{1} } u_{k_4\pi_1, \gamma_{1}})
\nonumber\\
&&
~~~~~~~~~~~~~~~~~~~~~~~~~~~~~
-(\delta_{-+++}) (u_{k_1\pi_1, \gamma_{1} }v_{k_4\pi_1,\gamma_{1} })
-(\delta_{-++-}) (u_{k_1\pi_1, \gamma_{1}}u_{k_4\pi_1,\gamma_{1} })
\big]
\nonumber\\
&&- (v_{k_2\pi_2, \gamma_{2}}^2) ~\times~
\big[
(\delta_{----}) (u_{k_1\pi_1, \gamma_{1} }u_{k_4\pi_1,\gamma_{1} })
+(\delta_{---+}) (u_{k_1\pi_1, \gamma_{1}}v_{k_4\pi_1,\gamma_{1} })
\nonumber\\
&&
~~~~~~~~~~~~~~~~~~~~~~~~~
+(\delta_{+---}) (v_{k_1\pi_1, \gamma_{1} }u_{k_4\pi_1,\gamma_{1} })
+(\delta_{+--+}) (v_{k_1\pi_1, \gamma_{1} }v_{k_4\pi_1, \gamma_{1} })
\big]
\nonumber\\
&&+ (u_{k_2\pi_2, \gamma_{2}} v_{k_2\pi_2, \gamma_{2}}) ~\times~
\big[
(\delta_{++--}) (v_{k_1\pi_1,\gamma_{1}}u_{k_4\pi_1,\gamma_{1} })
+(\delta_{++-+}) (v_{k_1\pi_1, \gamma_{1} }v_{k_4\pi_1,\gamma_{1} })
\nonumber\\
&&
~~~~~~~~~~~~~~~~~~~~~~~~~~~~~~~~~
+(\delta_{-+--}) (u_{k_1\pi_1, \gamma_{1} }u_{k_4\pi_1,\gamma_{1} })
+(\delta_{-+-+}) (u_{k_1\pi_1, \gamma_{1} }v_{k_4\pi_1, \gamma_{1} })
\big]
\nonumber\\
&&+ (v_{k_2\pi_2, \gamma_{2}} u_{k_2\pi_2, \gamma_{2}}) ~\times~
\big[
(\delta_{--++}) (u_{k_1\pi_1, \gamma_{1}}v_{k_4\pi_1,\gamma_{1} })
+(\delta_{--+-}) (u_{k_1\pi_1, \gamma_{1}}u_{k_4\pi_1,\gamma_{1} })
\nonumber\\
&&
~~~~~~~~~~~~~~~~~~~~~~~~~~~~~~~~~
+(\delta_{+-++}) (v_{k_1\pi_1, \gamma_{1} }v_{k_4\pi_1,\gamma_{1} })
+(\delta_{+-+-}) (v_{k_1\pi_1, \gamma_{1} }u_{k_4\pi_1,  \gamma_{1} })
\big]
\big\}~,  \nonumber\\
\eeqa
coefficient in front of $B^\dag_{k_3} B^{k_2}$
\beqa
&&h_{11}(k_i,\pi_i,\gamma_i )\nonumber\\
&&=\frac{1}{2} ~
\sum_{L}  \sum_{\lambda_i}
~\left(\frac{1}{2}\right)\frac{\sqrt{8(2L+1)}}{9} \frac{   (-1)^{L+J_2-J_1} }{ 2J_1+1 }
\bigg\{ \bigg\}
( \delta_{\pi_2 \pi_3}) (\delta_{k_1 k_4}\delta_{\pi_1 \pi_4})
( \delta_{J_1 J_4}\delta_{J_2 J_3} )
( \delta_{T_2 T_1}\delta_{T_4 T_3} \delta_{T_1  T_4})
( \delta_{Y_2 Y_1}\delta_{Y_4 Y_3} \delta_{Y_1  Y_4})\nonumber\\
&&\times
\big\{ (u_{k_1\pi_1, \gamma_{1}}^2) ~\times~
\big[
-(\delta_{----}) (v_{k_2\pi_3, \gamma_{3} }v_{k_3\pi_3, \gamma_{3} })
+(\delta_{--+-}) (v_{k_2\pi_3, \gamma_{3} }u_{k_3\pi_3,\gamma_{3} })
\nonumber\\
&&
~~~~~~~~~~~~~~~~~~~~~~~~~~~~~
+(\delta_{-+--}) (u_{k_2\pi_3,\gamma_{3} } v_{k_3\pi_3, \gamma_{3} })
-(\delta_{-++-}) (u_{k_2\pi_3, \gamma_{3} }u_{k_3\pi_3,\gamma_{3} })
\big]
\nonumber\\
&&- (v_{k_1\pi_1, \gamma_{1}}^2) ~\times~
\big[
(\delta_{++++}) (u_{k_2\pi_3, \gamma_{3} }u_{k_3\pi_3,\gamma_{3} })
-(\delta_{++-+}) (u_{k_2\pi_3, \gamma_{3} }v_{k_3\pi_3,\gamma_{3} })
\nonumber\\
&&
~~~~~~~~~~~~~~~~~~~~~~~~~
-(\delta_{+-++}) (v_{k_2\pi_3, \gamma_{3} }u_{k_3\pi_3,\gamma_{3} })
+(\delta_{+--+}) (v_{k_2\pi_3, \gamma_{3} }v_{k_3\pi_3,\gamma_{3} })
\big]
\nonumber\\
&&+ (u_{k_1\pi_1, \gamma_{1}} v_{k_1\pi_1, \gamma_{1}}) ~\times~
\big[
(\delta_{--++}) (v_{k_2\pi_3, \gamma_{3}}u_{k_3\pi_3,\gamma_{3} })
-(\delta_{---+}) (v_{k_2\pi_3, \gamma_{3}}v_{k_3\pi_3,\gamma_{3} })
\nonumber\\
&&
~~~~~~~~~~~~~~~~~~~~~~~~~~~~~~~~~
-(\delta_{-+++}) (u_{k_2\pi_3, \gamma_{3} }u_{k_3\pi_3,\gamma_{3} })
+(\delta_{-+-+}) (u_{k_2\pi_3, \gamma_{3} }v_{k_3\pi_3, \gamma_{3} })
\big]
\nonumber\\
&&+ (v_{k_1\pi_1, \gamma_{1}} u_{k_1\pi_1, \gamma_{1}}) ~\times~
\big[
(\delta_{++--}) (u_{k_2\pi_3, \gamma_{3}}v_{k_3\pi_3,\gamma_{3} })
-(\delta_{+++-}) (u_{k_2\pi_3, \gamma_{3}}u_{k_3\pi_3,\gamma_{3} })
\nonumber\\
&&
~~~~~~~~~~~~~~~~~~~~~~~~~~~~
-(\delta_{+---}) (v_{k_2\pi_3, \gamma_{3} }v_{k_3\pi_3,\gamma_{3} })
+(\delta_{+-+-}) (v_{k_2\pi_3, \gamma_{3} }u_{k_3\pi_3, \gamma_{3}})
\big]
\big\}~,\nonumber\\
\eeqa
coefficient in front of $ D^{\dag  k_2} D_{k_3}$
\beqa
&&h_{11}(k_i,\pi_i,\gamma_i )\nonumber\\
&&=\frac{1}{2} ~
\sum_{L}  \sum_{\lambda_i}
~\left(\frac{1}{2}\right)\frac{\sqrt{8(2L+1)}}{9} \frac{   (-1)^{L+J_2-J_1} }{ 2J_1+1 }
\bigg\{ \bigg\}
( \delta_{\pi_2 \pi_3}) (\delta_{k_1 k_4}\delta_{\pi_1 \pi_4})
( \delta_{J_1 J_4}\delta_{J_2 J_3} )
( \delta_{T_2 T_1}\delta_{T_4 T_3} \delta_{T_1  T_4})
( \delta_{Y_2 Y_1}\delta_{Y_4 Y_3} \delta_{Y_1  Y_4})\nonumber\\
&&\times
\big\{ (u_{k_1\pi_1, \gamma_{1}}^2) ~\times~
\big[
(\delta_{----}) (u_{k_2\pi_2, \gamma_{2} }u_{k_3\pi_2,\gamma_{2} })
+(\delta_{--+-}) (u_{k_2\pi_2, \gamma_{2} }v_{k_3\pi_2,\gamma_{2} })
\nonumber\\
&&
~~~~~~~~~~~~~~~~~~~~~~~~~~~~~
+(\delta_{-+--}) (v_{k_2\pi_2, \gamma_{2} }u_{k_3\pi_2,\gamma_{2} })
+(\delta_{-++-}) (v_{k_2\pi_2, \gamma_{2} }v_{k_3\pi_2,\gamma_{2} })
\big]
\nonumber\\
&&- (v_{k_1\pi_1, \gamma_{1}}^2) ~\times~
\big[
-(\delta_{++++}) (v_{k_2\pi_2, \gamma_{2} }v_{k_3\pi_2, \gamma_{2} })
-(\delta_{++-+}) (v_{k_2\pi_2, \gamma_{2} }u_{k_3\pi_2,\gamma_{2} })
\nonumber\\
&&
~~~~~~~~~~~~~~~~~~~~~~~~~
-(\delta_{+-++}) (u_{k_2\pi_2, \gamma_{2} }v_{k_3\pi_2,\gamma_{2} })
-(\delta_{+--+}) (u_{k_2\pi_2, \gamma_{2} }u_{k_3\pi_2,\gamma_{2} })
\big]
\nonumber\\
&&+ (u_{k_1\pi_1, \gamma_{1}} v_{k_1\pi_1, \gamma_{1}}) ~\times~
\big[
(\delta_{--++}) (u_{k_2\pi_2, \gamma_{2}}v_{k_3\pi_2,\gamma_{2} })
+(\delta_{---+}) (u_{k_2\pi_2, \gamma_{2}}u_{k_3\pi_2,\gamma_{2} })
\nonumber\\
&&
~~~~~~~~~~~~~~~~~~~~~~~~~~~~~~~~~
+(\delta_{-+++}) (v_{k_2\pi_2, \gamma_{2} }v_{k_3\pi_2,\gamma_{2} })
+(\delta_{-+-+}) (v_{k_2\pi_2, \gamma_{2} }u_{k_3\pi_2, \gamma_{2} })
\big]
\nonumber\\
&&+ (v_{k_1\pi_1, \gamma_{1}} u_{k_1\pi_1, \gamma_{1}}) ~\times~
\big[
(\delta_{++--}) (v_{k_2\pi_2, \gamma_{2}}u_{k_3\pi_2,\gamma_{2} })
+(\delta_{+++-}) (v_{k_2\pi_2, \gamma_{2} }v_{k_3\pi_2,\gamma_{2} })
\nonumber\\
&&
~~~~~~~~~~~~~~~~~~~~~~~~~~~~
+(\delta_{+---}) (u_{k_2\pi_2, \gamma_{2} }u_{k_3\pi_2,\gamma_{2} })
+(\delta_{+-+-}) (u_{k_2\pi_2, \gamma_{2} }v_{k_3\pi_2, \gamma_{2} })
\big]
\big\}.\nonumber\\
 \eeqa

Finally, the terms which appear in $\hat{H}_{20}+\hat{H}_{02}$, Eq. (\ref{H20}),  are
written as: the coefficient in front of $B^\dag_{k_1}D^{\dag k_4}$
\beqa
&&h_{20}(k_i,\pi_i,\gamma_i)\nonumber\\
&&=\frac{1}{2} ~
\sum_{L}\sum_{\lambda_i}
~\left(\frac{1}{2}\right)\frac{\sqrt{8(2L+1)}}{9} \frac{   (-1)^{L+J_2-J_1} }{ 2J_1+1 }
\bigg\{ \bigg\}
( \delta_{\pi_1 \pi_4}) (\delta_{k_2 k_3}\delta_{\pi_2 \pi_3})
( \delta_{J_2 J_3}\delta_{J_1 J_4} )
( \delta_{T_1 T_2}\delta_{T_2 T_3} \delta_{T_3  T_4})
( \delta_{Y_1 Y_2}\delta_{Y_2 Y_3} \delta_{Y_3  Y_4})\nonumber\\
&&\times
\big\{ (u_{k_2\pi_2, \gamma_{2}}^2) ~\times~
\big[
(\delta_{++++}) (u_{k_1\pi_1, \gamma_{q} }v_{k_4\pi_1,\gamma_{1} })
+(\delta_{+++-}) (u_{k_1\pi_1, \gamma_{q} }u_{k_4\pi_1,\gamma_{1} })
\nonumber\\
&&
~~~~~~~~~~~~~~~~~~~~~~~~~~~~~
-(\delta_{-+++}) (v_{k_1\pi_1, \gamma_{q} }v_{k_4\pi_1,\gamma_{1} })
-(\delta_{-++-}) (v_{k_1\pi_1, \gamma_{q} }u_{k_4\pi_1, \gamma_{1} })
\big]
\nonumber\\
%
%
&&- (v_{k_2\pi_2, \gamma_{2}}^2) ~\times~
\big[
+(\delta_{----}) (v_{k_1\pi_1, \gamma_{1} }u_{k_4\pi_1, \gamma_{1} })
+(\delta_{---+}) (v_{k_1\pi_1, \gamma_{1} }v_{k_4\pi_1,\gamma_{1} })
\nonumber\\
&&
~~~~~~~~~~~~~~~~~~~~~~~~~
-(\delta_{+---}) (u_{k_1\pi_1, \gamma_{1} }u_{k_4\pi_1,\gamma_{1} })
-(\delta_{+--+}) (u_{k_1\pi_1, \gamma_{1} }v_{k_4\pi_1,\gamma_{1} })
\big]
\nonumber\\
&&+ (u_{k_2\pi_2, \gamma_{2}} v_{k_2\pi_2, \gamma_{2}}) ~\times~
\big[
-(\delta_{++--}) (u_{k_1\pi_1,  \gamma_{1}}u_{k_4\pi_1,\gamma_{1} })
-(\delta_{++-+}) (u_{k_1\pi_1, \gamma_{1}}v_{k_4\pi_1,\gamma_{1} })
\nonumber\\
&&
~~~~~~~~~~~~~~~~~~~~~~~~~~~~~~~~~
+(\delta_{-+--}) (v_{k_1\pi_1, \gamma_{1} }u_{k_4\pi_1,\gamma_{1} })
+(\delta_{-+-+}) (v_{k_1\pi_1, \gamma_{1} }v_{k_4\pi_1,  \gamma_{1} })
\big]
\nonumber\\
&&+ (v_{k_2\pi_2, \gamma_{2}} u_{k_2\pi_2, \gamma_{2}}) ~\times~
\big[
(\delta_{--++}) (v_{k_1\pi_1,  \gamma_{1}}v_{k_4\pi_1,\gamma_{1} })
+(\delta_{--+-}) (v_{k_1\pi_1, \gamma_{1} }u_{k_4\pi_1,\gamma_{1} })
\nonumber\\
&&
~~~~~~~~~~~~~~~~~~~~~~~~~~~~
-(\delta_{+-++}) (u_{k_1\pi_1, \gamma_{1} }v_{k_4\pi_1,\gamma_{1} })
-(\delta_{+-+-}) (u_{k_1\pi_1, \gamma_{1} }u_{k_4\pi_1,  \gamma_{1} })
\big]
\big\}~,\nonumber\\
\eeqa
coefficient in front of $ D_{k_1} B^{k_4}$
\beqa
&&h_{02}(k_i,\pi_i,\gamma_i) \nonumber\\
&&=\frac{1}{2} ~
\sum_{L}\sum_{\lambda_i}
~\left(\frac{1}{2}\right)\frac{\sqrt{8(2L+1)}}{9} \frac{   (-1)^{L+J_2-J_1} }{ 2J_1+1 }
\bigg\{ \bigg\}
( \delta_{\pi_1 \pi_4}) (\delta_{k_2 k_3}\delta_{\pi_2 \pi_3})
( \delta_{J_2 J_3}\delta_{J_1 J_4} )
( \delta_{T_1 T_2}\delta_{T_2 T_3} \delta_{T_3  T_4})
( \delta_{Y_1 Y_2}\delta_{Y_2 Y_3} \delta_{Y_3  Y_4})\nonumber\\
&&\times
\big\{ (u_{k_2\pi_2, \gamma_{2}}^2) ~\times~
\big[
(\delta_{++++}) (v_{k_1\pi_1, \gamma_{1} }u_{k_4\pi_1,\gamma_{1} })
-(\delta_{+++-}) (v_{k_1\pi_1, \gamma_{1} }v_{k_4\pi_1,\gamma_{1}})
\nonumber\\
&&
~~~~~~~~~~~~~~~~~~~~~~~~~~~~~
+(\delta_{-+++}) (u_{k_1\pi_1, \gamma_{1} }u_{k_4\pi_1,\gamma_{1} })
-(\delta_{-++-}) (u_{k_1\pi_1, \gamma_{1} }v_{k_4\pi_1,\gamma_{1} })
\big]
\nonumber\\
&&- (v_{k_2\pi_2, \gamma_{2}}^2) ~\times~
\big[
+(\delta_{----}) (u_{k_1\pi_1, \gamma_{1} }v_{k_4\pi_1, \gamma_{1} })
-(\delta_{---+}) (u_{k_1\pi_1, \gamma_{1} }u_{k_4\pi_1,\gamma_{1} })
\nonumber\\
&&
~~~~~~~~~~~~~~~~~~~~~~~~~
+(\delta_{+---}) (v_{k_1\pi_1, \gamma_{1} }v_{k_4\pi_1,\gamma_{1} })
-(\delta_{+--+}) (v_{k_1\pi_1, \gamma_{1} }u_{k_4\pi_1, \gamma_{1} })
\big]
\nonumber\\
&&+ (u_{k_2\pi_2, \gamma_{2}} v_{k_2\pi_2, \gamma_{2}}) ~\times~
\big[
(\delta_{++--}) (v_{k_1\pi_1,  \gamma_{1}}v_{k_4\pi_1,\gamma_{1} })
-(\delta_{++-+}) (v_{k_1\pi_1, \gamma_{1} }u_{k_4\pi_1,\gamma_{1}})
\nonumber\\
&&
~~~~~~~~~~~~~~~~~~~~~~~~~~~~~~~~~
+(\delta_{-+--}) (u_{k_1\pi_1, \gamma_{1} }v_{k_4\pi_1,\gamma_{1} })
-(\delta_{-+-+}) (u_{k_1\pi_1, \gamma_{1} }u_{k_4\pi_1, \gamma_{1} })
\big]
\nonumber\\
&&+ (v_{k_2\pi_2, \gamma_{2}} u_{k_2\pi_2, \gamma_{2}}) ~\times~
\big[
-(\delta_{--++}) (u_{k_1\pi_1, \gamma_{1}}u_{k_4\pi_1,\gamma_{1} })
+(\delta_{--+-}) (u_{k_1\pi_1, \gamma_{1} }v_{k_4\pi_1,\gamma_{1} })
\nonumber\\
&&
~~~~~~~~~~~~~~~~~~~~~~~~~~~~
-(\delta_{+-++}) (v_{k_1\pi_1, \gamma_{1} }u_{k_4\pi_1,\gamma_{1} })
+(\delta_{+-+-}) (v_{k_1\pi_1, \gamma_{1} }v_{k_4\pi_1, \gamma_{1} })
\big]
\big\}~,\nonumber\\
\eeqa
coefficient in front of $ B^\dag_{k_3} D^{\dag k_2}$
\beqa
&&h_{20}(k_i,\pi_i,\gamma_i )\nonumber\\
&&=\frac{1}{2} ~
\sum_{L}  \sum_{\lambda_i}
~\left(\frac{1}{2}\right)\frac{\sqrt{8(2L+1)}}{9} \frac{   (-1)^{L+J_2-J_1} }{ 2J_1+1 }
\bigg\{ \bigg\}
( \delta_{\pi_2 \pi_3}) (\delta_{k_1 k_4}\delta_{\pi_1 \pi_4})
( \delta_{J_1 J_4}\delta_{J_2 J_3} )
( \delta_{T_2 T_1}\delta_{T_4 T_3} \delta_{T_1  T_4})
( \delta_{Y_2 Y_1}\delta_{Y_4 Y_3} \delta_{Y_1  Y_4})\nonumber\\
&&\times
\big\{ (u_{k_1\pi_1, \gamma_{1}}^2) ~\times~
\big[
+(\delta_{----}) (u_{k_2\pi_3, \gamma_{3} }v_{k_3\pi_3, \gamma_{3} })
-(\delta_{--+-}) (u_{k_2\pi_3, \gamma_{3}}u_{k_3\pi_3,\gamma_{3} })
\nonumber\\
&&
~~~~~~~~~~~~~~~~~~~~~~~~~~~~~
+(\delta_{-+--}) (v_{k_2\pi_3, \gamma_{3} }v_{k_3\pi_3,\gamma_{3} })
-(\delta_{-++-}) (v_{k_2\pi_3, \gamma_{3} }u_{k_3\pi_3, \gamma_{3} })
\big]
\nonumber\\
&&- (|v_{k_1\pi_1, \gamma_{q_1}}|^2) ~\times~
\big[
(\delta_{++++}) (v_{k_2\pi_3, \gamma_{3} }u_{k_3\pi_3,\gamma_{3} })
-(\delta_{++-+}) (v_{k_2\pi_3, \gamma_{3} }v_{k_3\pi_3,\gamma_{3} })
\nonumber\\
&&
~~~~~~~~~~~~~~~~~~~~~~~~~
+(\delta_{+-++}) (u_{k_2\pi_3, \gamma_{3} }u_{k_3\pi_3,\gamma_{3} })
-(\delta_{+--+}) (u_{k_2\pi_3, \gamma_{3} }v_{k_3\pi_3,\gamma_{3} })
\big]
\nonumber\\
&&+ (u_{k_1\pi_1, \gamma_{1}} v_{k_1\pi_1, \gamma_{1}}) ~\times~
\big[
-(\delta_{--++}) (u_{k_2\pi_3, \gamma_{3}}u_{k_3\pi_3,\gamma_{3} })
+(\delta_{---+}) (u_{k_2\pi_3, \gamma_{3} }v_{k_3\pi_3,\gamma_{3} })
\nonumber\\
&&
~~~~~~~~~~~~~~~~~~~~~~~~~~~~~~~~~
-(\delta_{-+++}) (v_{k_2\pi_3, \gamma_{3} }u_{k_3\pi_3,\gamma_{3} })
+(\delta_{-+-+}) (v_{k_2\pi_3, \gamma_{3} }v_{k_3\pi_3, \gamma_{3} })
\big]
\nonumber\\
&&+ (v_{k_1\pi_1, \gamma_{1}} u_{k_1\pi_1, \gamma_{1}}) ~\times~
\big[
(\delta_{++--}) (v_{k_2\pi_3, \gamma_{3}}v_{k_3\pi_3,\gamma_{3} })
-(\delta_{+++-}) (v_{k_2\pi_3, \gamma_{3} }u_{k_3\pi_3,\gamma_{3} })
\nonumber\\
&&
~~~~~~~~~~~~~~~~~~~~~~~~~~~~
+(\delta_{+---}) (u_{k_2\pi_3, \gamma_{3} }v_{k_3\pi_3,\gamma_{3} })
-(\delta_{+-+-}) (u_{k_2\pi_3, \gamma_{3} }u_{k_3\pi_3,\gamma_{3} })
\big]
\big\}~,\nonumber\\
\eeqa
coefficient in front of $ D_{k3} B^{k_2}$
\beqa
&&h_{02}(k_i,\pi_i,\gamma_i) \nonumber\\
&&=\frac{1}{2} ~
\sum_{L}  \sum_{\lambda_i}
~\left(\frac{1}{2}\right)\frac{\sqrt{8(2L+1)}}{9} \frac{   (-1)^{L+J_2-J_1} }{ 2J_1+1 }
\bigg\{ \bigg\}
( \delta_{\pi_2 \pi_3}) (\delta_{k_1 k_4}\delta_{\pi_1 \pi_4})
( \delta_{J_1 J_4}\delta_{J_2 J_3} )
( \delta_{T_2 T_1}\delta_{T_4 T_3} \delta_{T_1  T_4})
( \delta_{Y_2 Y_1}\delta_{Y_4 Y_3} \delta_{Y_1  Y_4})\nonumber\\
&&\times
\big\{ (u_{k_1\pi_1, \gamma_{1}}^2) ~\times~
\big[
+(\delta_{----}) (v_{k_2\pi_3, \gamma_{3} }u_{k_3\pi_3, \gamma_{3} })
+(\delta_{--+-}) (v_{k_2\pi_3, \gamma_{3} }v_{k_3\pi_3,\gamma_{3} })
\nonumber\\
&&
~~~~~~~~~~~~~~~~~~~~~~~~~~~~~
-(\delta_{-+--}) (u_{k_2\pi_3, \gamma_{3} }u_{k_3\pi_3,\gamma_{3} })
-(\delta_{-++-}) (u_{k_2\pi_3, \gamma_{3} }v_{k_3\pi_3,\gamma_{3} })
\big]
\nonumber\\
&&- (v_{k_1\pi_1, \gamma_{1}}^2) ~\times~
\big[
(\delta_{++++}) (u_{k_2\pi_3, \gamma_{3} }v_{k_3\pi_3,\gamma_{3} })
+(\delta_{++-+}) (u_{k_2\pi_3,\gamma_{3} }u_{k_3\pi_3,\gamma_{3} })
\nonumber\\
&&
~~~~~~~~~~~~~~~~~~~~~~~~~
-(\delta_{+-++}) (v_{k_2\pi_3, \gamma_{3} }v_{k_3\pi_3,\gamma_{3} })
-(\delta_{+--+}) (v_{k_2\pi_3, \gamma_{3} }u_{k_3\pi_3,\gamma_{3} })
\big]
\nonumber\\
&&+ (u_{k_1\pi_1, \gamma_{1}} v_{k_1\pi_1, \gamma_{1}}) ~\times~
\big[
(\delta_{--++}) (v_{k_2\pi_3, \gamma_{3}}v_{k_3\pi_3,\gamma_{3} })
+(\delta_{---+}) (v_{k_2\pi_3, \gamma_{3} }u_{k_3\pi_3,\gamma_{3} })
\nonumber\\
&&
~~~~~~~~~~~~~~~~~~~~~~~~~~~~~~~~~
-(\delta_{-+++}) (u_{k_2\pi_3, \gamma_{3} }v_{k_3\pi_3,\gamma_{3} })
-(\delta_{-+-+}) (u_{k_2\pi_3, \gamma_{3} }u_{k_3\pi_3,\gamma_{3} })
\big]
\nonumber\\
&&+ (v_{k_1\pi_1, \gamma_{1}} u_{k_1\pi_1, \gamma_{1}}) ~\times~
\big[
-(\delta_{++--}) (u_{k_2\pi_3, \gamma_{3}}u_{k_3\pi_3,\gamma_{3} })
-(\delta_{+++-}) (u_{k_2\pi_3, \gamma_{3} }v_{k_3\pi_3,\gamma_{3} })
\nonumber\\
&&
~~~~~~~~~~~~~~~~~~~~~~~~~~~~~~~~~
+(\delta_{+---}) (v_{k_2\pi_3, \gamma_{3} }u_{k_3\pi_3,\gamma_{3} })
+(\delta_{+-+-}) (v_{k_2\pi_3, \gamma_{3} }v_{k_3\pi_3, \gamma_{3} })
\big]
\big\}.\nonumber\\
\eeqa

In the above equations the $\bigg\{ \bigg\}$ is the same in Eq. (\ref{Ap-brackets}).

These matrix elements contain different 
combinations of $\delta_{\pm \pm \pm \pm}$-terms (see Eq. (\ref{Ap-deltas})), and 
the  product of the transformation coefficients
$\prod_i^4 \alpha^{J_i,T_i}_{\tau_i(N_il_i),\lambda_i,\pi_i,k_i}$ of
Eq. (\ref{Ap-brackets}).
Therefore, the matrix elements retain the information of the effective quark and antiquark
degrees of freedom, as well as the associated symmetry properties. Because the
prediagonalization requires a numerical procedure, which depends on the value of  $N_{cut}$,  
{\it i.e.} the dimension of the configurational space,
we have verified the stability of the results by changing $N_{cut}$ in
the range $N_{cut} \le 11$.

The numerical analysis shows that the following relations hold
\beqa
\delta_{1234}&=&\delta_{4321}~,\nonumber\\
\eeqa
and in the case of the substitution of quark for antiquark
\beqa
\delta_{++++}= \delta_{----} \nonumber\\
\delta_{-++-}= \delta_{+--+} 
\eeqa
are exact symmetries.

However, when changing quark for antiquark but with the same parity,
the following relations
\beqa
\delta_{++++}\approx \delta_{----} 
\eeqa
and 
\beqa
\delta_{-++-}\approx \delta_{+--+} 
\eeqa
are approximately fulfilled with a maximal deviation of the order of $4\%$ with respect to average
value $\frac{\delta_{++++}+\delta_{----}}{2} $ and
$\frac{\delta_{-++-}+\delta_{+--+}}{2} $, respectively.
This last symmetry is not a particle-antiparticle symmetry, because
for that also the parity has to be changed.
Nevertheless, these deviations
decreases as the maximal number of quanta $N_{cut}$
increases. In some cases the symmetry is restored up to $0.1\%$.


\begin{thebibliography}{99}

\bibitem{Weinberg} S. Weinberg, {\it The Quantum Theory of Fields}
  (Vol. II, Cambridge University Press, 1996).


\bibitem{Lee-book} T. D. Lee, {\it Particle Physics and Introduction
to Field Theory} (Harwood Academic Publishers, New York, 1981).


\bibitem{SO4-1} T. Yepez-Martinez, O. Civitarese and P. O. Hess, 
Int. J. Mod. Phys. E {\bf 25}, 1650067  (2016).

\bibitem{SO4-2} T. Yepez-Martinez, O. Civitarese and P. O. Hess, 
Int. J. Mod. Phys. E {\bf 26}, 1750012 (2017).

\bibitem{SO4-3} 
U. I. Ramirez-Soto, O. A. Rico-Trejo, T. Y\'epez-Mart\'inez,
P. O. Hess, A. Weber and O. Civitarese, 
J. Phys. G: Nucl. Part. Phys. {\bf 48}, 085013 (2021).






\bibitem{ChristLee} N. H. Christ and T. D. Lee, Phys. Rev. D {\bf 22}, 939 (1980).

\bibitem{Adam1996} A. Szczepaniak, E. S. Swanson, C. R. Ji and S. R.  Cotanch, Phys. Rev. Lett. {\bf 76}, 2011 (1996).

\bibitem{Adam2001} A. P. Szczepaniak and E. Swanson, Phys. Rev. D {\bf  65},  025012 (2001).


\bibitem{Hugo2004} C. Feuchter and H. Reinhardt, Phys. Rev. D {\bf    70}, 105021  (2004).

\bibitem{Hugo2011} H. Reinhardt, D. R. Campagnari and  A. P. Szczepaniak,  Phys. Rev. D {\bf 84}, 045006 (2011).


\bibitem{Yepez2012} T. Y\'epez-Mart\'inez, A. P. Szczepaniak and  H. Reinhardt, Phys. Rev. D {\bf 86}, 076010 (2012).


\bibitem{Greensite2015}  J. Greensite and A. P. Szczepaniak,  Phys. Rev. D 91, (2015) 034503.

\bibitem{Greensite2016} J. Greensite and A. P. Szczepaniak, Phys. Rev. D 93,  (2016) 074506.


\bibitem{Ring} P. Ring and P. Schuck, {\it The Nuclear Many Body   Problem} (Springer, Heidelberg, 1980).

\bibitem{Fetter} A. L. Fetter and J. D. Walecka, {\it Quantum Theory   of Many-Particle Systems} (Dover, New York, 2003).





\bibitem{Hess2006} P. O. Hess and A. P. Szczepaniak, Phys. Rev. C {\bf 73},  025201 (2006).

\bibitem{Yepez2010}  T. Y\'epez-Mart\ii nez, P. O. Hess,  A. P. Szczepaniak and O. Civitarese, Phys. Rev. C {\bf 81}, 045204 (2010).

\bibitem{Arturo2017}  D. A. Amor-Quiroz, T. Y\'epez-Mart\ii nez,  P. O. Hess, O. Civitarese,  and A. Weber,  Int. J. Mod. Phys. E  {\bf 26}, 1750082 (2017)






\bibitem{Bicudo2016} P. Bicudo, M. Cardoso, F. J. Llanes-Estrada and  T. V. Cauteren, Phys. Rev. D   {\bf 94}, 054006  (2016).






\bibitem{Finger}  J. R. Finger and J. E. Mandula, Nucl. Phys. B{\bf 199}, 168 (1982)

\bibitem{Adler} S. L. Adler and A. C. Davis, Nucl. Phys. B{\bf 244}, 469 (1984)

\bibitem{Yaouanc} A. Le Yaouanc, L. Oliver, O. P\`ene, and  J.-C. Raynal, Phys. Rev. D {\bf 29}, 1233 (1984)


\bibitem{Bicudo1990} P. J. de A. Bicudo and J. E. F. T. Ribeiro, Phys. Rev. D   {\bf 42}, 1611  (1990).

\bibitem{Llanes2000}   F. J. Llanes-Estrada and S. R. Cotanch,  Phys. Rev. Lett. {\bf 84}, 1102 (2000).

\bibitem{Llanes2002} F. J. Llanes-Estrada and S. R. Cotanch,  Nucl. Phys. A   {\bf 697}, 303  (2002).

\bibitem{Nefediev2008} A. V. Nefediev, J. E. F. T. Ribeiro, and  A. P. Szczepaniak, JETP Lett. {\bf 87}, 271 (2008).












\bibitem{Zwanziger2003} D. Zwanziger, Phys. Rev. Lett. {\bf 90},  102001 (2003).

\bibitem{Greensite2003}  J. Greensite and S. Olejnı´k,
  Phys. Rev. D {\bf 67}, 094503 ~2003!

\bibitem{Peng2008}
P. Guo, A. P. Szczepaniak, G. Galat'a, A. Vassallo, and E. Santopinto,
Phys. Rev. D {\bf 78}, 056003 (2008)


\bibitem{50yearsBCS}
Edited by R. A. Broglia and V. Zelevinsky,  Fifty years of nuclear BCS (World Scientific, 2020).

\bibitem{Bes}
D.R.Bes and G. G. Dussel, Nucl. Phys.  A {\bf 135} 1 (1969).

\bibitem{TDA-RPA-2021} T. Y\'epez-Mart\'inez, P. O. Hess and
  O. Civitarese, work in progress to be published.

\bibitem{PDG2020} P.A. Zyla et al. (Particle Data Group),  Prog. Theor. Exp. Phys. {\bf 2020}, 083C01 (2020). 

\end{thebibliography}
\end{document}